\documentclass[aip,jcp,numerical,reprint]{revtex4-1}
\usepackage[english]{babel}
\usepackage{microtype}
\usepackage{amsmath,amsthm,amssymb,amsfonts}
\usepackage{mathtools}
\usepackage{siunitx}
\usepackage{graphicx}
\usepackage{booktabs}
\usepackage{xr}
\usepackage[breaklinks=true,colorlinks=true,linkcolor=blue,urlcolor=blue,citecolor=blue]{hyperref}

\usepackage[bordercolor=none,backgroundcolor=none,textsize=scriptsize]{todonotes}

\newcommand{\ddf}[1]{\mathrm{d} #1 \,}
\DeclarePairedDelimiter\avg{\langle}{\rangle}

\DeclareMathOperator*{\argmin}{arg\,min}
\newcommand{\e}{\mathrm{e}}
\newcommand{\erf}{\operatorname{erf}}

\newcommand{\Ns}{N_{\text{s}}}

\usepackage{mhchem}

\DeclareSIUnit\molar{\textsc{M}}

\newcommand{\var}{\operatorname{var}}
\newcommand{\cov}{\operatorname{cov}}
\newcommand{\msd}{\operatorname{MSD}}

\begin{document}

\preprint{}

\title{Optimal estimates of self-diffusion coefficients from molecular dynamics simulations}

\author{Jakob T\'{o}mas Bullerjahn}
\author{S\"{o}ren von B\"{u}low}
\affiliation{Department of Theoretical Biophysics, Max Planck Institute of Biophysics, 60438 Frankfurt am Main, Germany}
\author{Gerhard~Hummer}
\email{gerhard.hummer@biophys.mpg.de}
\affiliation{Department of Theoretical Biophysics, Max Planck Institute of Biophysics, 60438 Frankfurt am Main, Germany}
\affiliation{Institute of Biophysics, Goethe University Frankfurt, 60438 Frankfurt am Main, Germany}
\date{\today}

\begin{abstract}
Translational diffusion coefficients are routinely estimated from molecular dynamics simulations.  Linear fits to mean squared displacement (MSD) curves have become the \emph{de facto} standard, from simple liquids to complex biomacromolecules.  Nonlinearities in MSD curves at short times are handled with a wide variety of \emph{ad hoc} practices, such as partial and piece-wise fitting of the data.  Here, we present a rigorous framework to obtain reliable estimates of the self-diffusion coefficient and its statistical uncertainty.  We also assess in a quantitative manner if the observed dynamics is indeed diffusive.  By accounting for correlations between MSD values at different times, we reduce the statistical uncertainty of the estimator and thereby increase its efficiency.  With a Kolmogorov-Smirnov test, we check for possible anomalous diffusion.  We provide an easy-to-use Python data analysis script for the estimation of self-diffusion coefficients.  As an illustration, we apply the formalism to molecular dynamics simulation data of pure TIP4P-D water and a single ubiquitin protein.  In a companion paper [J.~Chem.~Phys.~\textbf{XXX}, YYYYY (2020)], we demonstrate its ability to recognize deviations from regular diffusion caused by systematic errors in a common trajectory ``unwrapping'' scheme that is implemented in popular simulation and visualization software.  
\end{abstract}

%\pacs{87.15.Fh}
%\keywords{Suggested keywords}%Use showkeys class option if keyword display is desired

\maketitle

\section{Introduction}

Brownian motion is one of the pillars of biological physics, being observed on both microscopic and mesoscopic scales.  Einstein's seminal work\cite{Einstein1905} established the mean squared displacement (MSD) as the central observable to characterize the jittering motion of microscopic objects.  On sufficiently long time scales, the MSD of a freely diffusing tracer particle or macromolecule grows linearly in time with a slope directly proportional to its self-diffusion coefficient $D$.  Accordingly, $D$ is commonly estimated via linear fits to measured MSDs,\cite{QianSheetz1991} which at first glance may seem like a rigorous approach, because the resulting estimate is unbiased.  However, the precision of the estimate suffers if too many MSD values are used for the fit.\cite{WieserSchutz2008, Michalet2010}  This counter-intuitive behavior results from the fact that the most common estimator for the MSD of a finite time series $\{ X_{0}, X_{1}, \dots, X_{N-1}, X_{N} \}$, namely
\begin{equation}\label{eq:msd-data}
\msd_{i} = \sum_{n=0}^{N-i} \frac{(X_{n+i} - X_{n})^{2}}{N-i+1} \, ,
\end{equation}
introduces correlations between the $\msd_{i}$ at different time lags $t_i=i\Delta t$ with $i = 1, 2, \dots, M \leq N$.  Furthermore, the underlying dynamics may not be purely diffusive, as is often the case with molecular dynamics (MD) simulations.  Non-diffusive dynamics, arising, \emph{e.g.},~due to ballistic motion,\cite{HuangChavez2011} short-lived memory\cite{LiBian2015} or caging effects,\cite{vanMegenSchope2017} affect the MSD values at short times.  As an example of a complex diffusion process where the transition to regular diffusion is slow, Vargas and Snurr\cite{VargasSnurr2015} analyzed the translational diffusion of alkanes in an anisotropic metal-organic framework.  A common strategy to characterize non-diffusive short-time dynamics is to invoke elaborate non-Markov processes,\cite{LysyPillai2016} but these have to be specifically tailored to the data at hand.  We therefore face the problem that diffusion coefficient estimates using short-time data are compromised by possible non-diffusive dynamics, and estimates using long-time data suffer from large statistical uncertainties.  

To address the latter, various proposals for improved diffusion coefficient estimation have been published in recent years,\cite{Berglund2010, Michalet2010, MichaletBerglund2012, VestergaardBlainey2014} correcting for experimental artifacts and making more efficient use of the available data.  Here, we introduce a rigorous framework that combines these sophisticated estimators with a sub-sampling procedure, where the length of the recording-time interval is varied to suppress nonlinearities in the MSD curves on short time scales.  The resulting sub-sampled dynamics is then modeled via a diffusive process $X$ combined with an instantaneous stepwise spread, satisfying
\begin{equation}\label{eq:msd-model}
\avg{\msd_{i}} \equiv \avg{( X_{n+i} - X_{n})^{2}} = a^{2} + i \sigma^{2} \, ,
\end{equation}
whose details are presented at the beginning of Sec.~\ref{sec:theory}.  The associated self-diffusion coefficient is given by
\begin{equation*}
D = \frac{\sigma^{2}}{2 \Delta t} \, ,
\end{equation*}
where $\Delta t$ denotes the length of the time interval between two consecutive observations $X_{i}$ and $X_{i+1}$ $\forall i$.  By introducing the static noise parameter $a^{2}$ into our generic model of diffusion, we aim to account for  ``molecular events'' such as correlated collisions and cage diffusion without having to use complex system-specific dynamical models.  A price we pay is that the process $X$ becomes valid only for time intervals $\Delta t$ that exceed the time scale of the molecular events.  We note that ballistic motion, \emph{e.g.}, in an underdamped Langevin equation would lead to $a^{2} < 0$.  However, in practice, such down-shifts are normally not observed, because they get compensated by other effects that broaden the MSD.  

The process $X$ is compatible with the above-mentioned diffusion coefficient estimators, a few of which we briefly review and then compare performance-wise in Sec.~\ref{sec:theory}.  To determine the range of interval lengths where $X$ appropriately describes the data at hand, we propose a quality factor based on $\chi^{2}$-statistics in Sec.~\ref{subsec:optimal-timestep} to decide whether an estimate overfits or underfits the model to the data.  As a probe of possible anomalous diffusion, the resulting diffusion coefficient estimate obtained at short times can then be compared to observed long-time dynamics, as described in Sec.~\ref{subsec:kolmogorov-smirnov-test}.  We illustrate our framework by applying it to various MD trajectories in Sec.~\ref{sec:results}: First on an ensemble of pure TIP4P-D water\cite{PianaDonchev2015} (Secs.~\ref{subsec:TIP4P-D-simulation-setup}--\ref{subsec:TIP4P-D-diffusion-coefficient}) and then on a single ubiquitin protein solvated in TIP4P-D water (Sec.~\ref{subsec:ubiquitin-diffusion-coefficient}).  Although the analysis of the latter system reveals some ambiguities, our framework identifies the dynamics as diffusive with a self-diffusion coefficient that also properly captures ubiquitin's long-time behavior (Sec.~\ref{sec:verifying-predictions}).  In the companion paper, Ref.~\onlinecite{vonBuelowBullerjahn2020}, we show that unreasonable quality-factor values arise due to previously overlooked shortcomings in a trajectory ``unwrapping'' scheme used widely in the analysis of MD simulations at constant pressure.  Finally, Sec.~\ref{sec:conclusions} provides a summary of our results and the Appendix gives the interested reader further details on some of the more elaborate derivations.

\section{Diffusion coefficient estimation}\label{sec:theory}

At short times, inertial dynamics and correlated local motions lead to deviations from simple diffusion typically resulting in a fast initial spread of the particle position.  On longer time scales, the resulting MSD then resembles Eq.~\eqref{eq:msd-model}.  Let us therefore consider two discrete-time Wiener processes, $Z$ and $X$, where the latter process evolves on top of each realization $Z_{i}$ of the former.  The dynamics of the two processes is captured by the following iterative equations,
\begin{subequations}\label{eqs:Z_X_processes}
\begin{align}
Z_{i+1} & = Z_{i} + \sigma R_{i} \, , & & \avg{R_{i}} = 0 \, , & & \avg{R_{i} R_{j}} = \delta_{i,j} \, ,
\\
X_{i} & = Z_{i} + \frac{a}{\sqrt{2}} S_{i} \, , & & \avg{S_{i}} = 0 \, , & & \avg{S_{i} S_{j}} = \delta_{i,j} \, ,
\end{align}
\end{subequations}
where $R$ and $S$ denote uncorrelated normal distributed random variables with zero mean and unit variance, and $\delta_{i,j}$ is the Kronecker delta that evaluates to one if $i=j$ and zero otherwise.  By construction, we have $\avg{Z_{i} S_{j}} = 0$ $\forall i,j$ and $\avg{Z_{i} R_{j}} = 0$ $\forall i \leq j$.  Because $Z$ and $X$ are also both normal distributed, their distributions are fully characterized by the mean and (co)variance, namely
\begin{align*}
\avg{Z_{i}} & = 0 \, , & & \cov(Z_{i},Z_{j}) = \sigma^{2} \min(i,j) \, ,
\\
\avg{X_{i}} & = \avg{Z_{i}} = 0 \, , & & \cov(X_{i},X_{j}) = \sigma^{2} \min(i,j) + \frac{a^{2}}{2} \delta_{i,j} \, ,
\end{align*}
which result in Eq.~\eqref{eq:msd-model} for the MSD of $X$.  

In the following, we shall discuss a few established diffusion coefficient estimators within the context of the process $X$, and then compare them performance-wise.

\subsection{Ordinary least squares estimators}\label{subsec:ols-estimators}

A linear fit to the MSD corresponds to minimizing the sum of squared residuals between the data [Eq.~\eqref{eq:msd-data}] and the model [Eq.~\eqref{eq:msd-model}], thus giving rise to a set of \emph{ordinary least squares} (OLS) estimators
\begin{equation}\label{eq:ols}
(a_{\text{OLS}}^{2},\sigma_{\text{OLS}}^{2}) = \argmin_{a^{2}, \sigma^{2} \geq 0} \sum_{i=1}^{M} (\msd_{i} - a^{2} - i \sigma^{2})^{2}
\end{equation}
for some $M \leq N$.  In non-pathological cases, the OLS problem [Eq.~\eqref{eq:ols}] is analytically tractable and results in the following expressions for the estimators,
\begin{align}\label{eqs:ols-estimators}
a_{\text{OLS}}^{2} & = \frac{\beta \gamma - \alpha \delta}{M \beta - \alpha^{2}} \, , & \sigma_{\text{OLS}}^{2} & = \frac{M \delta - \alpha \gamma}{M \beta - \alpha^{2}} \, ,
\\
\notag
\alpha & = \frac{M(M+1)}{2} \, , & \beta & = \alpha \frac{2M + 1}{3} \, , 
\\
\notag
\gamma & = \sum_{i = 1}^{M} \msd_{i} \, , & \delta & = \sum_{i = 1}^{M} i \, \msd_{i} \, .  
\end{align}
Although these estimators are unbiased, their precision drastically decreases for increased numbers $M$ of MSD values used in the fit.  This counter-intuitive behavior can be read off their corresponding variances (see Appendix~\ref{app:OLS-variances}) and, to circumvent this shortcoming, it was suggested in Ref.~\onlinecite{Michalet2010} to vary the value of $M$ to single out the estimators with the smallest variances.  The associated standard deviation is typically much larger than \emph{ad hoc} uncertainty estimates, such as those constructed from fits to hand-selected linear regimes in the data.\cite{AbrahamvanderSpoel2019}

\subsection{Covariance-based estimators}\label{subsec:cve-estimators}

In Ref.~\onlinecite{VestergaardBlainey2014}, an alternative to the OLS estimators was proposed, which circumvents the computation of $\msd_{i}$ via Eq.~\eqref{eq:msd-data} altogether.  It makes use of the fact that
\begin{align*}
\avg{(X_{n+1} - X_{n})(X_{n} - X_{n-1})} = - \frac{a^{2}}{2}
\end{align*}
must hold $\forall n$, as well as Eq.~\eqref{eq:msd-model} evaluated at $i=1$.  This results in the unbiased covariance-based estimators (CVE) 
\begin{subequations}\label{eqs:cve-estimators}
\begin{align}
a_{\text{CVE}}^{2} & = - 2 \sum_{n=1}^{N-1} \frac{(X_{n+1} - X_{n})(X_{n} - X_{n-1})}{N-1} \, , 
\\
\sigma_{\text{CVE}}^{2} & = \sum_{n=0}^{N-1} \frac{(X_{n+1} - X_{n})^{2}}{N} - a_{\text{CVE}}^{2} \, ,
\end{align}
\end{subequations}
whose variances are given by
\begin{subequations}\label{eqs:cve-variances}
\begin{align}
\var(a_{\text{CVE}}^{2})
& = \frac{ 7 a^{4} + 8 a^{2} \sigma^{2} + 4 \sigma^{4} }{(N - 1)} - \frac{ 2 a^{4} }{(N - 1)^{2}} \, ,
\\
\var(\sigma_{\text{CVE}}^{2})
\notag
& = 4 \frac{a^{2} \sigma^{2} + \sigma^{4}}{N-1} + 2 \frac{a^{4} + \sigma^{4}}{N} + \frac{5 a^{4} + 4 a^{2} \sigma^{2}}{N (N-1)}
\\
\label{eq:cve-sigma-variance}
& \mathrel{\phantom{=}} - \frac{a^{4}}{(N-1)^{2}} - \frac{a^{4}}{N^{2} (N-1)^{2}} \, .  
\end{align}
\end{subequations}
While Eqs.~\eqref{eqs:cve-estimators} and~\eqref{eqs:cve-variances} are computationally inexpensive to evaluate, they are only guaranteed to be practically optimal for signal-to-noise ratios $\sigma^{2} / a^{2}$ larger than one.\cite{VestergaardBlainey2014}

\subsection{Generalized least squares estimators}\label{subsec:gls-estimators}

To improve upon the OLS scheme, one can take into account correlations between the residuals, which are collected in the covariance matrix of the $\msd_{i}$-values with elements (see Appendix~\ref{app:derivation-1})
\begin{subequations}\label{eqs:covariance_matrices}
\begin{align}\label{eq:general_covariance_matrix}
\Sigma_{i,j}(a^{2},\sigma^{2})
\notag
& = \Sigma_{i,j}(0,\sigma^{2}) + \frac{a^{4}(1 + \delta_{i,j}) + 4 a^{2} \sigma^{2} \min(i,j)}{N - \min(i,j) + 1}
\\
& \mathrel{\phantom{=}} + \frac{a^{4} \max(0,N-i-j+1)}{(N - i + 1) (N - j + 1)}
\end{align}
for $i,j = 1, 2, \dots, M$ and
\begin{align}\label{eq:specific_covariance_matrix}
\notag
\Sigma_{i,j} & (0,\sigma^{2}) = \frac{\sigma^{4}}{3} \bigg[ \frac{2 \min(i,j) \big[ 1 + 3 i j - \min(i,j)^{2} \big]}{N - \min(i,j) + 1}
\\
\notag
& \mathrel{\phantom{=}} + \frac{\min(i,j)^{2} - \min(i,j)^{4}}{(N-i+1)(N-j+1)} + \Theta(i+j-N-2)
\\
& \mathrel{\phantom{=}} \times \frac{(N+1-i-j)^{4} - (N+1-i-j)^{2}}{(N-i+1)(N-j+1)} \bigg] \, ,
\end{align}
\end{subequations}
where $\Theta(z)$ denotes the Heaviside unit step function.  Equation~\eqref{eq:ols} is then replaced by
\begin{subequations}\label{eqs:gls}
\begin{align}
(a_{\text{GLS}}^{2},\sigma_{\text{GLS}}^{2})
& = \argmin_{a^{2}, \sigma^{2} \geq 0} \chi^{2}(a^{2},\sigma^{2}) \, ,
\\
\chi^{2}(a^{2},\sigma^{2})
\notag
& = \sum_{i,j = 1}^{M} (\msd_{i} - a^{2} - i \sigma^{2}) \Sigma^{-1}_{i,j}(a_{\text{GLS}}^{2},\sigma_{\text{GLS}}^{2})
\\
\label{eq:chi-squared}
& \mathrel{\phantom{=}} \times (\msd_{j} - a^{2} - j \sigma^{2}) \, , 
\end{align}
\end{subequations}
in a procedure commonly referred to as \emph{generalized least squares} (GLS).  Note that the covariance matrix is evaluated at $(\smash{a^{2}},\smash{\sigma^{2}}) = (\smash{a_{\text{GLS}}^{2}}, \smash{\sigma_{\text{GLS}}^{2}})$, and thus remains fixed while $\smash{a^{2}}$ and $\smash{\sigma^{2}}$ are being varied.  The \emph{a priori} unknown $\smash{a_{\text{GLS}}^{2}}$ and $\smash{\sigma_{\text{GLS}}^{2}}$ are then found by requiring self-consistency.  

In general, there exist no closed-form expressions for the estimators of the GLS problem [Eq.~\eqref{eqs:gls}].  They satisfy the following transcendental coupled equations,
\begin{align}\label{eqs:gls-estimators}
a_{\text{GLS}}^{2} & = \frac{\mu \nu - \lambda \xi}{\kappa \mu - \lambda^{2}} \, , \qquad \sigma_{\text{GLS}}^{2} = \frac{\kappa \xi - \lambda \nu}{\kappa \mu - \lambda^{2}} \, , 
\\
\notag
\kappa & = \sum_{i,j = 1}^{M} \Sigma_{i,j}^{-1}(a_{\text{GLS}}^{2},\sigma_{\text{GLS}}^{2}) \, , 
\\
\notag
\lambda & = \sum_{i,j = 1}^{M} i \, \Sigma_{i,j}^{-1}(a_{\text{GLS}}^{2},\sigma_{\text{GLS}}^{2}) \, , 
\\
\notag
\mu & = \sum_{i,j = 1}^{M} i \, j \, \Sigma_{i,j}^{-1}(a_{\text{GLS}}^{2},\sigma_{\text{GLS}}^{2}) \, , 
\\
\notag
\nu & = \sum_{i,j = 1}^{M} \msd_{i} \, \Sigma_{i,j}^{-1}(a_{\text{GLS}}^{2},\sigma_{\text{GLS}}^{2}) \, , 
\\
\notag
\xi & = \sum_{i,j = 1}^{M} i \, \msd_{j} \, \Sigma_{i,j}^{-1}(a_{\text{GLS}}^{2},\sigma_{\text{GLS}}^{2}) \, , 
\end{align}
which have to be solved numerically, \emph{e.g.}, via iterative root-finding methods.  In Appendix~\ref{app:algorithm}, we provide one such algorithm, based on fixed-point iteration, which we have implemented in a Python data analysis script.\cite{PythonScript}

Analogous to $\smash{a_{\text{OLS}}^{2}}$ and $\smash{\sigma_{\text{OLS}}^{2}}$, the estimators in Eqs.~\eqref{eqs:gls-estimators} are asymptotically unbiased.  Lower bounds for the variances of the GLS estimators can be inferred from the inverse of the Fisher information matrix $\mathbf{I}$ associated with the likelihood
\begin{equation*}
\mathcal{L}(a^{2},\sigma^{2}) \propto \exp \left( - \chi^{2}(a^{2},\sigma^{2}) / 2 \right) \, ,
\end{equation*}
which is constructed from the $\chi^{2}$-statistic [Eq.~\eqref{eq:chi-squared}].  In our case, the components of $\mathbf{I}$ read
\begin{align*}
I_{1,1} & = \frac{1}{2} \frac{\partial^{2} \chi^{2}(a^{2},\sigma^{2})}{(\partial a^{2})^{2}} = \kappa \, ,
\\
I_{1,2} \equiv I_{2,1} & = \frac{1}{2} \frac{\partial^{2} \chi^{2}(a^{2},\sigma^{2})}{\partial a^{2} \partial \sigma^{2}} = \lambda \, ,
\\
I_{2,2} & = \frac{1}{2} \frac{\partial^{2} \chi^{2}(a^{2},\sigma^{2})}{(\partial \sigma^{2})^{2}} = \mu \, ,
\end{align*}
and result in the inverse matrix
\begin{equation*}
\mathbf{I}^{-1} = \frac{1}{\kappa \mu - \lambda^{2}} \begin{pmatrix}
\mu & -\lambda
\\
-\lambda & \kappa
\end{pmatrix} \, .  
\end{equation*}
The variances of the GLS estimators are thus estimated from below by
\begin{align}\label{eqs:gls-variances}
\var(a_{\text{GLS}}^{2}) \geq \frac{\mu}{\kappa \mu - \lambda^{2}} \, , & & \var(\sigma_{\text{GLS}}^{2}) \geq \frac{\kappa}{\kappa \mu - \lambda^{2}} \, .  
\end{align}
Whenever the estimators are Gaussian, equality holds in Eqs.~\eqref{eqs:gls-variances}.

\subsection{Properties of the GLS estimators}

It can be shown (see Appendix~\ref{app:derivation-2}) that Eqs.~\eqref{eqs:gls-estimators} are of the form
\begin{subequations}\label{eq:asymptotic-gls-estimators}
\begin{align}
a_{\text{GLS}}^{2} & = \frac{\nu - \msd_{1} \lambda + \mathcal{O} \big( [ \sigma^{2} / a^{2} ]^{-2} \big)}{\kappa - \lambda + \mathcal{O} \big( [ \sigma^{2} / a^{2} ]^{-2} \big)} \, , 
\\
\sigma_{\text{GLS}}^{2} & = \frac{\msd_{1} \kappa - \nu + \mathcal{O} \big( [ \sigma^{2} / a^{2} ]^{-2} \big)}{\kappa - \lambda + \mathcal{O} \big( [ \sigma^{2} / a^{2} ]^{-2} \big)} \, ,
\end{align}
which implies, on the one hand, that
\begin{equation}\label{eq:asymptotic-estimator-sum}
a_{\text{GLS}}^{2} + \sigma_{\text{GLS}}^{2} \mathop{\sim}^{\sigma^{2} \gg a^{2}} \msd_{1}
\end{equation}
\end{subequations}
must hold for sufficiently large signal-to-noise ratios $\sigma^{2} / a^{2}$, and, on the other hand, that in the special case of $a \equiv 0$, the estimator becomes analytically tractable $\forall M,N$ and simply reads
\begin{equation}\label{eq:gls-estimator}
\sigma_{\text{GLS}}^{2} \vert_{a=0} \equiv \msd_{1} \, .  
\end{equation}
In this case, Eqs.~\eqref{eqs:gls-variances} reduce to
\begin{equation}\label{eq:gls-variance_a=0}
\var(\sigma_{\text{GLS}}^{2} \vert_{a=0}) \geq \mu^{-1} \vert_{a=0} = \frac{2}{N} \sigma^{4} \, .  
\end{equation}

Another special case for which Eq.~\eqref{eq:asymptotic-estimator-sum} becomes exact is given by $M=2$.  Equations~\eqref{eqs:gls-estimators} are then analytically tractable and give rise to the closed-form estimators
\begin{subequations}\label{eqs:M=2-estimators}
\begin{align}
a_{M=2}^{2} & = 2 \msd_{1} - \msd_{2} \, ,
\\
\sigma_{M=2}^{2} & = - \msd_{1} + \msd_{2} \, ,
\end{align}
\end{subequations}
with the following variances,
\begin{subequations}\label{eqs:M=2-variances}
\begin{align}
\var (a_{M=2}^{2})
& = 4 \Sigma_{1,1} - 4 \Sigma_{1,2} + \Sigma_{2,2} \, ,
\\
\var (\sigma_{M=2}^{2})
\label{eq:M=2-sigma-variance}
& = \Sigma_{1,1} - 2 \Sigma_{1,2} + \Sigma_{2,2} \, .  
\end{align}
\end{subequations}
These results coincide with the OLS estimators and their respective variances (given in Appendix~\ref{app:OLS-variances}) for $M=2$.

\subsection{Application to three-dimensional time series}  

Up until now, we have solely focused on one-dimensional time series, while two- and three-dimensional particle trajectories are recorded in experiments and MD simulations.  For a three-dimensional time series, we can decompose the associated MSD in the following way,
\begin{equation}\label{eq:3D-msd}
\msd_{i}^{\text{3D}} = \msd_{x,i} + \msd_{y,i} + \msd_{z,i} \, ,
\end{equation}
where the $\smash{\msd_{d,i}}$-values are obtained by evaluating Eq.~\eqref{eq:msd-data} using the respective one-dimensional time series along each spatial dimension $d \in \{x,y,z\}$.  If we furthermore model the stochastic dynamics along every Cartesian coordinate via our minimal diffusion process $X$, then the expectation value of Eq.~\eqref{eq:3D-msd} is simply given by
\begin{align*}
\avg{\msd_{i}^{\text{3D}}} & = a_{x}^{2} + a_{y}^{2} + a_{z}^{2} + i (\sigma_{x}^{2} + \sigma_{y}^{2} + \sigma_{z}^{2}) \, ,
\\
& \equiv a_{\text{3D}}^{2} + i \sigma_{\text{3D}}^{2} \, ,
\end{align*}
with $D = \sigma_{\text{3D}}^{2} / 6 \Delta t$.  The elements of the associated covariance matrix read
\begin{equation*}
\Sigma_{i,j}^{\text{3D}} = \Sigma_{i,j}(a_{x}^{2},\sigma_{x}^{2}) + \Sigma_{i,j}(a_{y}^{2},\sigma_{y}^{2}) + \Sigma_{i,j}(a_{z}^{2},\sigma_{z}^{2}) \, .  
\end{equation*}
Here, we have assumed no correlation between dimensions, \emph{i.e.}, $\smash{\avg{\msd_{d_{1},i} \msd_{d_{2},j}}} = \smash{\avg{\msd_{d_{1},i}}} \smash{\avg{\msd_{d_{2},j}}}$.  This linear behavior propagates all the way up to the estimators $\smash{\theta_{\text{EST}}^{2}} \in \{ \smash{a_{\text{EST}}^{2}}, \smash{\sigma_{\text{EST}}^{2}} \}$ for $\text{EST} \in \{ \text{OLS}, \text{GLS}, \text{CVE} \}$ and their respective variances, resulting in
\begin{align*}
\theta_{\text{3D},\text{EST}}^{2} & = \theta_{x,\text{EST}}^{2} + \theta_{y,\text{EST}}^{2} + \theta_{z,\text{EST}}^{2} \, ,
\\
\var(\theta_{\text{3D},\text{EST}}^{2}) & = \var(\theta_{x,\text{EST}}^{2}) + \var(\theta_{y,\text{EST}}^{2}) + \var(\theta_{z,\text{EST}}^{2}) \, .  
\end{align*}
The latter relation follows from the fact that $\smash{\avg{\theta_{d_{1},\text{EST}}^{2} \theta_{d_{2},\text{EST}}^{2}}} = \smash{\avg{\theta_{d_{1},\text{EST}}^{2}}} \smash{\avg{\theta_{d_{2},\text{EST}}^{2}}} = \smash{\theta_{d_{1}}^{2} \theta_{d_{2}}^{2}}$ holds for all the estimators considered in this paper.  Finally, a $\chi^{2}$-statistic can be calculated for three-dimensional time series by replacing $\msd_{i}$, $\smash{a^{2}}$, $\smash{\sigma^{2}}$ and $\smash{\Sigma_{i,j}(a_{\text{GLS}}^{2},\sigma_{\text{GLS}}^{2})}$ in Eq.~\eqref{eq:chi-squared} with their three-dimensional analogues, giving
\begin{align}\label{eq:3D-chi-squared}
\chi^{2}
\notag
& = 3 \sum_{i,j = 1}^{M} ( \msd_{i}^{\text{3D}} - a_{\text{3D}}^{2} - i \sigma_{\text{3D}}^{2}) \Sigma_{i,j}^{-1} (a_{\text{3D,GLS}}^{2},\sigma_{\text{3D,GLS}}^{2})
\\
& \mathrel{\phantom{=}} \times ( \msd_{j}^{\text{3D}} - a_{\text{3D}}^{2} - j \sigma_{\text{3D}}^{2}) \, .  
\end{align}

\subsection{Comparing the estimators}\label{subsec:estimator-comparison}

\begin{figure}[t!]
\begin{center}
\includegraphics{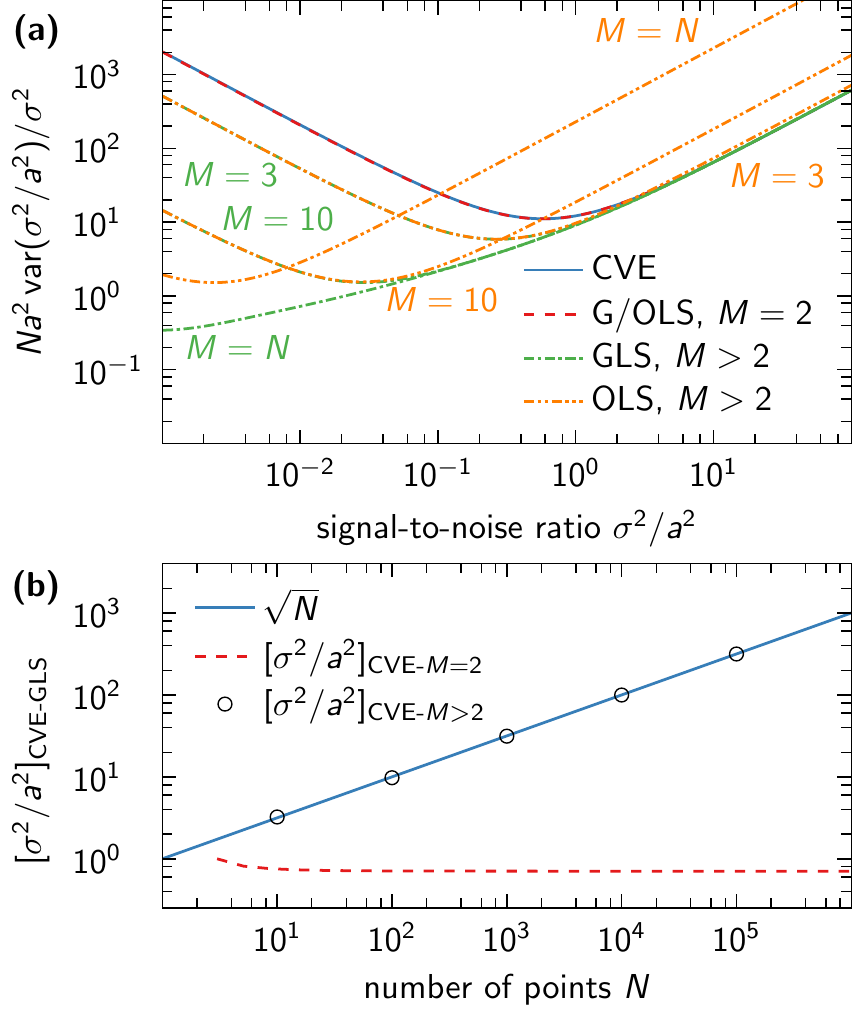}
\caption{Quantitative comparison of the estimators.  (a)~Estimator variances as functions of the signal-to-noise ratio for $N=100$.  For $\sigma^{2}/a^{2} > 1$, the CVE variance [Eq.~\eqref{eq:cve-sigma-variance}, solid blue line] is almost indistinguishable from the general GLS solution [Eq.~\eqref{eqs:gls-variances}, dash-dotted green lines] and the closed-form special solution for $M=2$ [Eq.~\eqref{eq:M=2-sigma-variance}, dashed red line], while the OLS variance [Eq.~\eqref{eq:ols-sigma-variance}, dash-double-dotted orange lines] gradually worsens with increasing $M$.  In turn, at low signal-to-noise ratios and $M \ll N$, the OLS estimators have comparable uncertainties to their GLS counterparts.  (b)~Signal-to-noise threshold below which the GLS estimator outperforms the CVE estimator as a function of the number of points $N$ in the time series.  Unlike $\smash{[ \sigma^{2} / a^{2} ]_{\text{CVE-}M=2}}$, given by Eq.~\eqref{eq:cve-M=2} (red dashed line), the CVE-GLS intersection for $M > 2$ is only numerically tractable (black open symbols).  Yet, it turns out that its $N$-dependence is, for sufficiently large $M$, remarkably well captured by a square root function (solid blue line).  }
\label{fig:variance_comparison}
\end{center}
\end{figure}

Because all the above-mentioned estimators are unbiased, we must compare their respective variances to rank them.  Under the assumption that all four estimator pairs predict roughly the same values for $a^{2}$ and $\sigma^{2}$, which is realized in practice for $N \gg 1$, we can compare their corresponding variances as functions of the signal-to-noise ratio $\sigma^{2} / a^{2}$.  In general, the GLS estimators outperform their counterparts at low and intermediate signal-to-noise ratios, as visualized in Fig.~\ref{fig:variance_comparison}a.  Only for $\sigma^{2} / a^{2} \gg 1$ do the CVE estimators eventually catch up, as seen by their intersection point with the closed-form GLS estimators.  For $M=2$, the estimator variances cross at
\begin{equation}\label{eq:cve-M=2}
\bigg[ \frac{\sigma^{2}}{a^{2}} \bigg]_{\text{CVE-}M=2} = \sqrt{\frac{N - 1}{2 N - 4}} \, .  
\end{equation}
For $M>2$, the crossing point must be found numerically and behaves roughly like (see Fig.~\ref{fig:variance_comparison}b)
\begin{equation*}
\bigg[ \frac{\sigma^{2}}{a^{2}} \bigg]_{\text{CVE-}M > 2} \approx \sqrt{N} \, .  
\end{equation*}
Therefore, for $M>2$, the GLS estimators tend to be superior already for comparably short trajectory lengths.

\section{Statistical testing of model assumptions and quality of fit}\label{sec:methods}

Another advantage of the GLS estimators is their natural compatibility with the quality factor that is introduced below and serves as a measure for the quality of diffusion coefficient estimates.  This section also provides a test to probe for possible anomalous diffusion in the long-time limit.

\subsection{Determining the optimal length of the recording-time interval}\label{subsec:optimal-timestep}

If non-diffusive short-time dynamics are present in a time series, their influence can be suppressed via sub-sampling, where intermediate observations from the original time series $\{ X_{0}, X_{1}, \dots, X_{N-1}, X_{N} \}$ are removed to generate a set of shorter time series
\begin{gather*}
\{ X_{0}, X_{1}, X_{2}, \dots \} \, , 
\\
\{ X_{0}, X_{2}, X_{4}, \dots \} \, , 
\\
\{ X_{0}, X_{3}, X_{6}, \dots \} \, ,
\\
\vdots
\\
\{ X_{0}, X_{n}, X_{2n}, \dots \} \, ,
\end{gather*}
with respective recording-time interval lengths $\Delta t_{1}, \Delta t_{2}, \Delta t_{3}, \dots, \Delta t_{n}$, where $\Delta t_{n} = n \Delta t_{1}$.  Alternatively, one can choose different starting points to sub-sample the time series; \emph{e.g.}, it is just as valid to use $\{ X_{1}, X_{3}, X_{5}, \dots \}$ instead of $\{ X_{0}, X_{2}, X_{4}, \dots \}$.  While a longer interval will generally result in a more linear MSD curve, at least for $n \ll N$, the shortened length of the new time series adversely affects the uncertainty of the diffusion coefficient estimate.  The estimator uncertainty can be improved somewhat by including in the analysis all $n-1$ interspersed time series constructed from the discarded points between $X_{i}$ and $X_{i+n}$.  However, these additional time series are not fully independent, so their inclusion has only a modest effect and is therefore not considered in what follows.  

We balance the competition between systematic and statistical errors at short and long $\Delta t_{n}$, respectively, by exploiting the fact that the GLS estimators originate from the $\smash{\chi^{2}}$-statistic in Eq.~\eqref{eq:chi-squared}.  For a sample of GLS estimates $\smash{\{ {a_{\mathrlap{\text{GLS}}}^{2}}^{(k)} \, ,{\sigma_{\mathrlap{\text{GLS}}}^{2}}^{(k)} \}_{k = 1, 2, \dots, \Ns}}$, the corresponding $\smash{\chi^{2}}$-values should follow a $\smash{\chi^{2}}$-distribution with $M-2$ degrees of freedom whenever the residuals
\begin{align}\label{eq:residuals}
\delta\msd_{i}^{(k)}
\notag
& = \sum_{j=1}^{M} \big( \msd_{j}^{(k)} - {a_{\mathrlap{\text{GLS}}}^{2}}^{(k)} - j {\sigma_{\mathrlap{\text{GLS}}}^{2}}^{(k)} \big)
\\
& \mathrel{\phantom{=}} \times \Sigma_{i,j}^{-1/2} \big( a_{\text{GLS}}^{2}, \sigma_{\text{GLS}}^{2} \big)
\end{align}
are normal distributed with $\smash{\Sigma_{i,j}^{-1/2}}$ denoting the elements of the inverse square root of the covariance matrix [Eq.~\eqref{eqs:covariance_matrices}] evaluated at $(\smash{a^{2}},\smash{\sigma^{2}}) = (\smash{a_{\text{GLS}}^{2}}, \smash{\sigma_{\text{GLS}}^{2}})$.  If there are nonlinearities present at short times that skew the distribution of the residuals, we expect atypical $\chi^{2}$-values and the recording-time interval has to be lengthened.  However, instead of focusing on the broadly distributed $\chi^{2}$-values, we instead consider the associated quality factor
\begin{equation}\label{eq:quality-factor}
Q(\Delta t_{n},M) = 1 - \frac{\gamma \left( M/2 - 1, \chi^{2} / 2 \right)}{\Gamma(M/2-1)} \, ,
\end{equation}
which coincides with the reciprocal cumulative distribution function (CDF) of the $\chi^{2}$-statistic and therefore only takes values between zero and one.  Here,
\begin{align*}
\gamma(a,z) = \int_{0}^{z} \ddf{x} x^{a-1} \e^{-x} \, , & & \Gamma(a) = \lim_{z \to \infty} \gamma(a,z) \, , 
\end{align*}
denote the lower incomplete and ordinary $\Gamma$-functions, respectively.  The quality factor can be seen as the probability to observe a $\smash{\chi^{2}}$-value greater than $\smash{\chi^{2}(a_{\text{GLS}}^{2},\sigma_{\text{GLS}}^{2})}$.  

Ideally, the $\smash{Q^{(k)}}$-values, computed from the estimates $\smash{{a_{\mathrlap{\text{GLS}}}^{2}}^{(k)}}$ and $\smash{{\sigma_{\mathrlap{\text{GLS}}}^{2}}^{(k)}}$, are distributed uniformly on $[0,1]$ with a sample average $\overline{Q} \approx 1/2$.  Here, $\smash{\overline{O}}$ denotes the arithmetic mean of a finite sample of observations $\smash{ \{ O^{(k)} \}_{k=1,2,\dots,\Ns}}$, namely
\begin{equation}\label{eq:sample-average}
\overline{O} = \frac{1}{\Ns} \sum_{k=1}^{\Ns} O^{(k)} \, .  
\end{equation}
If the average quality factor is significantly lower than $1/2$, certain features of the data are not well captured by our diffusive model, thus requiring us to lengthen the recording-time interval.  Conversely, $\smash{\overline{Q}} > 1/2$ hints at overfitting.  The optimal interval length $\Delta t_{\text{opt}}$ marks the instance, where the quality factor first reaches $\smash{\overline{Q}} \approx 1/2$ and the residuals in Eq.~\eqref{eq:residuals} follow a normal distribution.  If the time series behind the $\smash{Q^{(k)}}$-values are sufficiently long, then the quality factor should remain approximately constant well beyond $\Delta t_{\text{opt}}$, before slowly drifting off towards unity.  

As an alternative to the above procedure, one could use only parts of the data, \emph{e.g.}, the time series along $x$ and $y$, to evaluate the GLS estimators.  The quantities $\chi^{2}$ and $Q$ would then serve as tools to measure how well the estimators fit the remaining data.  Such use of the above concepts would only require minor alterations to Eqs.~\eqref{eq:chi-squared} and~\eqref{eq:quality-factor}, in particular changing the number of degrees of freedom from $M-2$ to $M$.

\subsection{Comparing short-time predictions to observed dynamics at long times}\label{subsec:kolmogorov-smirnov-test}

A common concern is that regular diffusion cannot fully account for the dynamics in complex molecular systems, which must then be described using more elaborate models that incorporate memory or trapping effects.  To make sure our short-time diffusion coefficient estimate $D(\Delta t_{n} = \Delta t_{\text{opt}})$ can correctly describe the actual dynamics of our system at long times, we take advantage of the fact that typical MD trajectories are long compared to the time range $M \Delta t_{\text{opt}}$ used to fit the diffusion coefficient.  We therefore compare the statistics of the relative endpoints $\smash{\Delta X^{(k)} = X_{N}^{(k)} - X_{0}^{(k)}}$ of each considered time series, characterized by the empirical (cumulative) distribution function (eCDF), to the CDF
\begin{equation}\label{eq:cdf}
F(\Delta X) = \frac{1}{2} + \frac{1}{2} \erf \bigg( \frac{\Delta X - \overline{\Delta X}}{\sqrt{2 a^{2}(\Delta t_{\text{opt}}) + 4 D(\Delta t_{\text{opt}}) N \Delta t_{1}}} \bigg)
\end{equation}
of a diffusive process using the Kolmogorov-Smirnov (KS) statistic
\begin{equation}\label{eq:kolmogorov-smirnov-statistic}
S = \max_{1 \leq k \leq \Ns} \bigg( \frac{k}{\Ns} - F(\Delta X^{(k)}) , F(\Delta X^{(k)}) - \frac{k-1}{\Ns} \bigg) \, .  
\end{equation}
Here, $\erf(z)$ denotes the error function.  The discrepancy test above measures the absolute size of the largest difference between the two distribution functions, and the test statistic $S$ can be used to compute a corresponding \textit{p}-value, \emph{i.e.}, the probability that, if the endpoints had actually been drawn from $F'(\Delta X)$, the resulting KS statistic would have been greater or equal to $S$.  Alternatively, one can vary $D$ in Eq.~\eqref{eq:cdf} to find the diffusion coefficient that minimizes Eq.~\eqref{eq:kolmogorov-smirnov-statistic} and thus best describes the observed long-time dynamics.  This is possible because $a^{2}(\Delta t_{\text{opt}}) \ll 2 D(\Delta t_{\text{opt}}) N \Delta t_{1}$ for $N \gg 1$, so $a^{2}$ can either be neglected or kept constant at $a^{2} = a^{2}(\Delta t_{\text{opt}})$ while $D$ is varied.  

In the following, we refer to the use of Eqs.~\eqref{eq:cdf} and~\eqref{eq:kolmogorov-smirnov-statistic} as a KS test.

\section{Results and discussion}\label{sec:results}

\subsection{MD simulation of TIP4P-D water}\label{subsec:TIP4P-D-simulation-setup}

We tested the applicability of the GLS estimator for the diffusion coefficient by performing a \SI{1}{\micro \second} MD simulation of water at ambient conditions.  We placed 4139 TIP4P-D water molecules\cite{PianaDonchev2015} in a cubic simulation box with an approximate edge length of \SI{5}{\nano \meter} and periodic boundary conditions.  We recorded the coordinates of each atom in the system at intervals of $\Delta t_{1} = \SI{1}{\pico \second}$ to capture possible non-diffusive short-time dynamics.  The simulation was run at $\SI{300}{\kelvin}$\cite{BussiDonadio2007} and $\SI{1}{\bar}$\cite{ParrinelloRahman1981} using Gromacs/2018.6\cite{AbrahamMurtola2015} with temperature- and pressure-coupling constants $\tau_{T} = \SI{1}{\pico \second}$ and $\tau_{p} = \SI{5}{\pico \second}$, respectively.  After an initial \SI{100}{\pico \second} equilibration period in the $NVT$-ensemble, and a consecutive \SI{5}{\nano \second} run in the $NpT$-ensemble, the production run ensued at the same temperature and pressure.  We made use of the leap-frog integrator with a \SI{2}{\femto \second} time step.  All atomic bonds to hydrogens were treated with the LINCS constraint algorithm.\cite{HessBekker1997}

For data evaluation, we made use of an improved scheme\cite{vonBuelowBullerjahn2020} to ``unwrap'' the three-dimensional center-of-mass trajectories of every water molecule out of the simulation box, which we then split into three independent time series -- one for every Cartesian coordinate.  These one-dimensional time series were used to estimate the MSD along the spatial dimensions $d \in \{x,y,z\}$ via Eq.~\eqref{eq:msd-data}, after which the estimates were summed up to give the total MSD for every molecule $k=1, 2, \dots, 4139$ according to Eq.~\eqref{eq:3D-msd}.  All trajectories displayed a clear non-diffusive initial regime, which became less pronounced as the length of the recording-time interval was increased, followed by a more-or-less linear growth at longer times.

\subsection{Establishing ground truth for TIP4P-D water}

\begin{figure*}[t!]
\begin{center}
\includegraphics{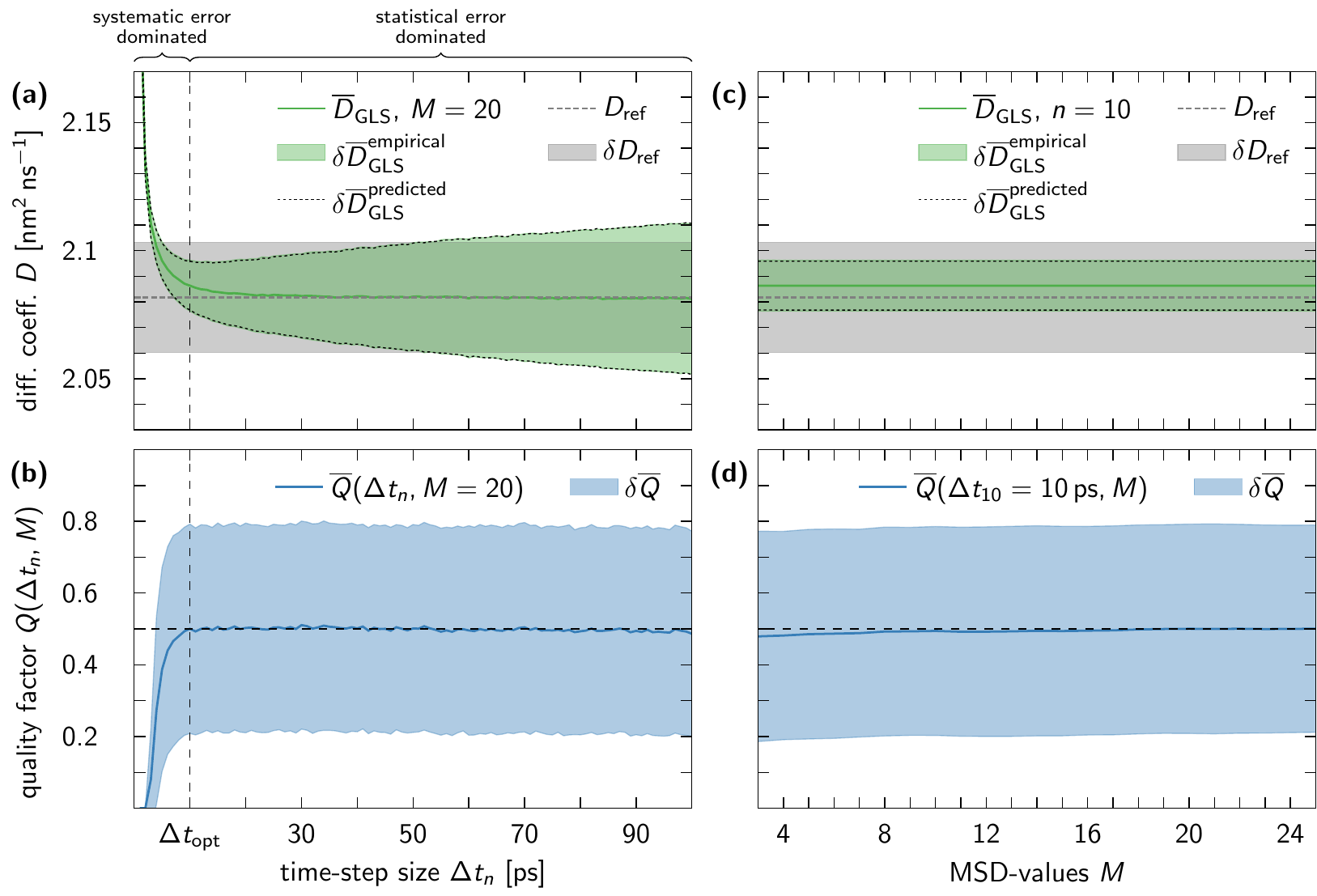}
\caption{Translational diffusion coefficient estimation and quality factor analysis of TIP4P-D water simulation data.  (a)~Using the GLS estimators [Eqs.~\eqref{eqs:gls-estimators}], in combination with Eqs.~\eqref{eqs:diffusion-coefficient-estimation}, we computed the average diffusion coefficient at different recording-time interval lengths $\Delta t_{n}$ with $M=20$ fixed (solid green line).  Our results are together with the ground-truth value $D_{\text{ref}} = \SI{2.08}{\nano \meter \squared \per \nano \second}$ (dashed gray line) and its uncertainty $\delta D_{\text{ref}} = \pm \SI{0.02}{\nano \meter \squared \per \nano \second}$ (gray shaded area) for comparison, which were determined at a fixed interval length of $\Delta t_{10} = \SI{10}{\pico \second}$ and therefore do not vary with $\Delta t_{n}$.  The uncertainty of the GLS estimates (green shaded area) was determined empirically via Eq.~\eqref{eq:D-std} and is nicely reproduced by our analytic prediction [Eq.~\eqref{eq:D-uncertainty-estimate}, dotted black lines] for virtually all interval lengths.  (b)~A lower bound $\Delta t_{\text{opt}}$ to the range of desirable interval lengths was determined from the instance, where the sample average of the quality factor [Eqs.~\eqref{eq:quality-factor} and~\eqref{eq:sample-average}, solid blue line] converges to $\smash{\overline{Q}} \approx 1/2$, namely at $\Delta t_{\text{opt}} = \SI{10}{\pico \second}$.  The shaded area represents one sample standard deviation of uncertainty.  (c--d)~Same as (a) and (b), respectively, except that here the length of the recording-time interval was kept fixed at $\Delta t_{\text{opt}} = \SI{10}{\pico \second}$, while $M$ was varied.  The estimated diffusion coefficients are independent of the choice of $M$.  
}
\label{fig:diffusion_coefficient_water}
\end{center}
\end{figure*}

To establish a reference value $D_{\text{ref}}$ for the diffusion coefficient to compare our results to, we analyzed a sub-sampled version of our time series with $\Delta t_{n} = \SI{10}{\pico \second}$, using the OLS estimators for $M=20$.  The length of the recording-time interval was chosen such that it was long enough to suppress non-diffusive short-time effects, while minimizing its impact on the length $N+1$ of the time series and, in turn, the variance.  For every molecule $k$, we obtained an estimate 
\begin{subequations}\label{eqs:diffusion-coefficient-estimation}
\begin{equation}
D_{d,\text{EST}}^{(k)} = \frac{{\sigma^{2}_{\mathrlap{d,\text{EST}}}}^{(k)} \ \;}{2 \Delta t_{n}}
\end{equation}
for the translational diffusion coefficient along each spatial dimension $d$, which we then used to calculate the sample averages $\smash{\overline{D}}_{d,\text{OLS}}$ via Eq.~\eqref{eq:sample-average} with $\Ns = 4139$.  Unsurprisingly, we found perfect agreement of the $\smash{\overline{D}}_{d,\text{OLS}}$-values [$\smash{\overline{D}}_{x,\text{OLS}} = \smash{\overline{D}}_{y,\text{OLS}} = \smash{\overline{D}}_{z,\text{OLS}} = \SI{2.08 \pm 0.04}{\nano \meter \squared \per \nano \second}$], thus confirming that the overall diffusive motion of the water molecules was isotropic with a scalar translational diffusion coefficient
\begin{equation}\label{eq:D-mean}
\overline{D}_{\text{EST}} = \frac{\overline{D}_{x,\text{EST}} + \overline{D}_{y,\text{EST}} + \overline{D}_{z,\text{EST}}}{3} \, ,
\end{equation}
whose uncertainty was determined from the sample standard deviation as follows,
\begin{align}\label{eq:D-std}
\notag
\delta & \overline{D}_{\text{EST}}^{\text{empirical}} = 
\\
& \frac{1}{3} \sqrt{\sum_{k=1}^{\Ns} \frac{\big( D_{x,\text{EST}}^{(k)} + D_{y,\text{EST}}^{(k)} + D_{z,\text{EST}}^{(k)} - 3 \overline{D}_{\text{EST}} \big)^{2}}{\Ns-1}} \, .  
\end{align}
\end{subequations}
Evaluating these expressions with the OLS estimates $\smash{\{ {a_{\mathrlap{d,\text{GLS}}}^{2}}^{(k)} \ \; ,{\sigma_{\mathrlap{d, \text{GLS}}}^{2}}^{(k)} \ \; \}_{k = 1, 2, \dots, 4139}^{d \in \{x, y, z\}}}$ finally resulted in the reference value ${D_{\text{ref}} = \SI{2.08 \pm 0.02}{\nano \meter \squared \per \nano \second}}$, which is well within the range of previously reported values for TIP4P-D water.\cite{PianaDonchev2015}  It is slightly lower than the experimentally determined value of $\SI{2.3}{\nano \meter \squared \per \nano \second}$, measured at $\SI{298}{\kelvin}$.\cite{KrynickiGreen1978}  The main reasons for this discrepancy are, on the one hand, that the TIP4P-D model slightly overestimates the viscosity of water\cite{vonBuelowSiggel2019} and, on the other hand, that finite-size effects in simulations using periodic boundary conditions further reduce the diffusion coefficient.  Although closed-form corrections have been developed both for three-dimensional in-bulk diffusion\cite{YehHummer2004} and for quasi two-dimensional diffusion within membranes,\cite{VoegeleKoefinger2018} we will not make use of them here, because finite-size effects are systematic and therefore not estimator-specific.  The in-bulk correction is, however, accounted for in our Python data analysis script.\cite{PythonScript}

\subsection{Diffusion coefficient estimation and quality factor analysis for TIP4P-D water}\label{subsec:TIP4P-D-diffusion-coefficient}

Analogous to the previous section, we considered the center-of-mass trajectories of every molecule $k$ along each spatial dimension $d$, for which we computed the associated MSDs via Eq.~\eqref{eq:msd-data}.  These were substituted into Eqs.~\eqref{eqs:gls-estimators} to obtain the GLS estimates $\smash{{a_{\mathrlap{d,\text{GLS}}}^{2}}^{(k)} \ \; }$ and $\smash{{\sigma_{\mathrlap{d,\text{GLS}}}^{2}}^{(k)} \ \; }$ that, in turn, were used to calculate the sample average and uncertainty of the diffusion coefficient via Eqs.~\eqref{eqs:diffusion-coefficient-estimation}.  This procedure was repeated for different sub-sampling interval lengths $\Delta t_{n}$ in the range of \SIrange{1}{100}{\pico \second}, where $\Delta t_{1} = \SI{1}{\pico \second}$ and $M=20$ were kept fixed.  Our results are summarized in Fig.~\ref{fig:diffusion_coefficient_water}a, where our GLS estimate for the diffusion coefficient, $\smash{\overline{D}_{\text{GLS}}}$, is plotted next to the reference value $D_{\text{ref}}$.  Figure~\ref{fig:diffusion_coefficient_water}a also compares the empirically estimated uncertainty [Eq.~\eqref{eq:D-std}] to the predicted uncertainty
\begin{equation}\label{eq:D-uncertainty-estimate}
\delta \overline{D}_{\text{GLS}}^{\text{prediction}} =  \sqrt{\ \; \sum_{\mathclap{d=x,y,z}} \frac{\var(\overline{\sigma}^{2}_{d,\text{GLS}})}{(6 \Delta t_{n})^{2}}} \, ,
\end{equation}
which was evaluated with the help of Eqs.~\eqref{eqs:gls-variances}.  The analytical and numerical estimates of the uncertainties, Eqs~\eqref{eq:D-std} and~\eqref{eq:D-uncertainty-estimate}, coincide near perfectly for most interval lengths.  

A lower bound $\Delta t_{\text{opt}}$ to the range of viable interval lengths, where the trade-off between estimate quality and uncertainty is balanced, was determined via the quality factor [Eq.~\eqref{eq:quality-factor}], which we evaluated for all molecules present in the system using different recording-time interval lengths $\Delta t_{n}$.  We thereby calculated a corresponding $\chi^{2}$-value for each molecule $k$ according to Eq.~\eqref{eq:3D-chi-squared}.  Figure~\ref{fig:diffusion_coefficient_water}b visualizes the sample average and uncertainty of our $Q^{(k)}$-values, where the latter was estimated via the sample standard deviation.  Because the sample average converged to $\overline{Q} \approx 1/2$ around $n = 10$ and did not vary significantly at higher $n$, the lower bound was chosen to be $\Delta t_{\text{opt}} = \SI{10}{\pico \second}$.  At this interval length, the average GLS estimate for the diffusion coefficient may not be fully converged, but $D_{\text{ref}}$ and $\overline{D}_{\text{GLS}}$ are clearly within each other's uncertainty intervals.  We verified that the chosen bound was independent of our choice of $M$ by repeating our analysis with $n = 10$ fixed and varying $M$ (see Figs.~\ref{fig:diffusion_coefficient_water}c--d).  Indeed, at the chosen interval length, both the diffusion coefficient estimate and the quality factor did not vary significantly with $M$.  

\begin{figure}[t!]
\begin{center}
\includegraphics{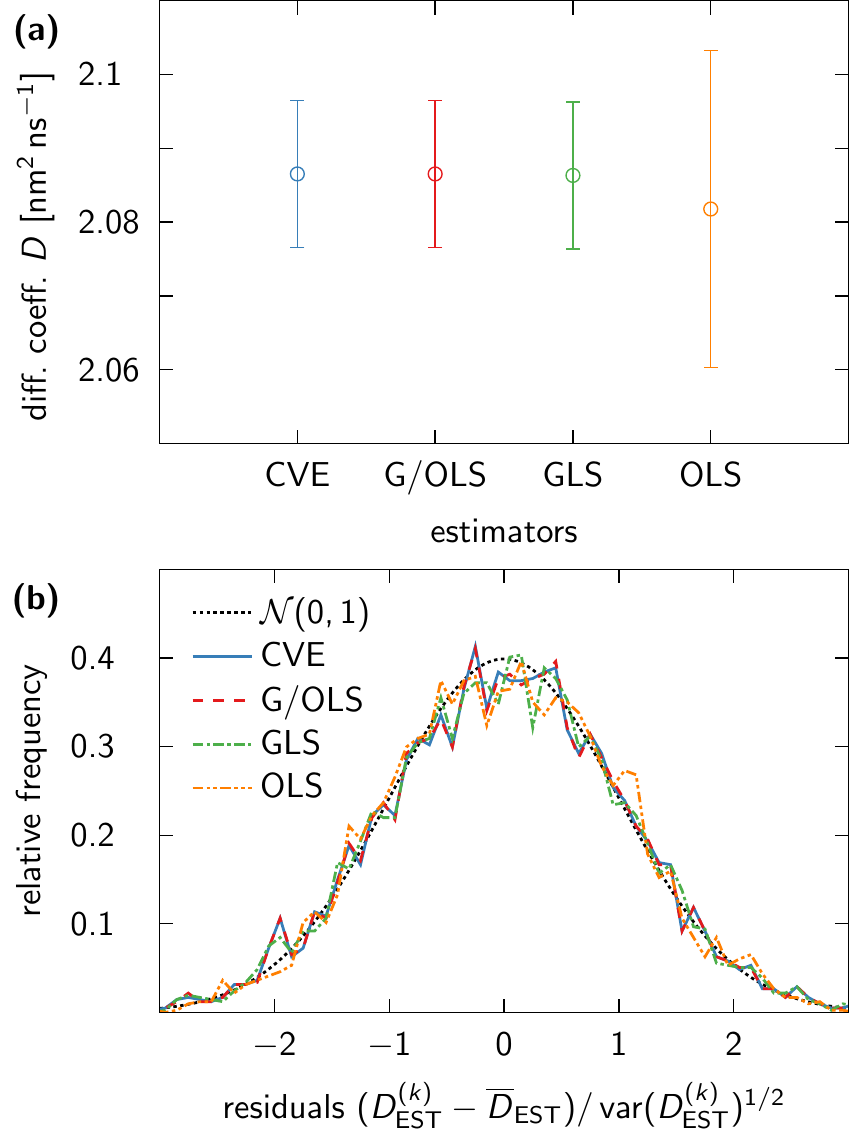}
\caption{Comparison of estimator performances when applied to TIP4P-D water simulation data, recorded at $\Delta t_{n} = \SI{10}{\pico \second}$, with $M=20$.  (a)~Diffusion coefficient predictions by the four estimators discussed in this paper.  For each water molecule, we evaluated the OLS , CVE , GLS  and G/OLS estimators [Eqs.~\eqref{eqs:ols-estimators}, \eqref{eqs:cve-estimators}, \eqref{eqs:gls-estimators} and~\eqref{eqs:M=2-estimators}, respectively], and substituted the results into Eqs.~\eqref{eqs:diffusion-coefficient-estimation} to calculate the sample mean $\smash{\overline{D}_{\text{EST}}}$ (open symbols) and standard deviation $\smash{\delta \overline{D}_{\text{EST}}^{\text{empirical}}}$ (error bars) for all $\text{EST} \in \{ \text{OLS}, \text{GLS}, \text{CVE} \}$, respectively.  (b)~Distribution of scaled residuals $\smash{(D^{(k)}_{\text{EST}} - \overline{D}_{\text{EST}}) / \var(D^{(k)}_{\text{EST}})^{1/2}}$.  The appropriately scaled residuals of all four estimators follow a standard normal distribution (black dotted line), meaning that our uncertainty estimates [Eq.~\eqref{eq:Dk-uncertainty-estimate}] are appropriate and that the assumptions underlying the use of the quality factor [Eq.~\eqref{eq:quality-factor}] are met.  
}
\label{fig:estimator_comparison_water}
\end{center}
\end{figure}

As discussed in the companion paper, Ref.~\onlinecite{vonBuelowBullerjahn2020}, we originally used the built-in tool \texttt{trjconv} in Gromacs to unwrap the simulation trajectories, which resulted in a highly oscillating diffusion coefficient estimate for short recording-time intervals, and a slower convergence of the quality factor to $\overline{Q} \approx 1/2$.  These irregularities intensified as the edge length of the simulation box was reduced, which made us wary of the unwrapping scheme implemented in \texttt{trjconv}, and prompted us to develop an improved unwrapping scheme appropriate for simulations at constant pressure.\cite{vonBuelowBullerjahn2020}

Finally, in Fig.~\ref{fig:estimator_comparison_water}a, we give a comparison of all the different estimators discussed in this paper, evaluated for our TIP4P-D water trajectories with $\smash{\Delta t_{n}} = \Delta t_{\text{opt}} = \SI{10}{\pico \second}$ and, when appropriate, $M=20$.  At this interval length, we estimated an average noise parameter of $\smash{\overline{a}_{\text{GLS}}^{2}} = \SI{8.4(5)e-3}{\nano \meter \squared}$ for TIP4P-D water, which gives a fairly high signal-to-noise ratio ($\sigma^{2} / a^{2} \approx 15$), where all estimators, aside from the OLS estimators, have essentially the same uncertainty (see Fig.~\ref{fig:variance_comparison}).  The ratio is still below the $[\sigma^{2}/a^{2}]_{\text{CVE-}M > 2}$-threshold, meaning that the GLS estimators outperform, if only marginally, the other three.  Figure~\ref{fig:estimator_comparison_water}b demonstrates that the scaled residuals $\smash{(D^{(k)}_{\text{EST}} - \overline{D}_{\text{EST}}) / \var(D^{(k)}_{\text{EST}})^{1/2}}$ follow a $\mathcal{N}(0,1)$-distribution, \emph{i.e.}, a Gaussian with zero mean and unit variance, for all four considered estimators.  Here, $\smash{D^{(k)}_{\text{EST}}} = \smash{D_{x,\text{EST}}^{(k)}} + \smash{D_{y,\text{EST}}^{(k)}} + \smash{D_{z,\text{EST}}^{(k)}}$, $\smash{\overline{D}_{\text{EST}}}$ is defined via Eq.~\eqref{eq:D-mean} and 
\begin{equation}\label{eq:Dk-uncertainty-estimate}
\var\big(D^{(k)}_{\text{EST}}\big) = \sum_{\mathclap{d=x,y,z}} \frac{\var\big({\sigma^{2}_{\mathrlap{d,\text{EST}}}}^{(k)} \ \; \big)}{(6 \Delta t_{n})^{2}}
\end{equation}
is evaluated using either Eqs.~\eqref{eqs:cve-variances}, \eqref{eqs:gls-variances}, \eqref{eqs:M=2-variances} or~\eqref{eqs:ols-variances}, depending on the estimator $\text{EST} \in \{ \text{OLS}, \text{GLS}, \text{CVE} \}$.  The good agreement with the standard normal distribution at the chosen interval length implies, on the one hand, that our predictions for the estimator-uncertainties are appropriate and, on the other hand, that the $\smash{{\sigma^{2}_{\mathrlap{\text{EST}}}}^{(k)}} = \smash{{\sigma^{2}_{\mathrlap{x,\text{EST}}}}^{(k)} \ \;} + \smash{{\sigma^{2}_{\mathrlap{y,\text{EST}}}}^{(k)} \ \;} + \smash{{\sigma^{2}_{\mathrlap{z,\text{EST}}}}^{(k)} \ \;}$ are all Gaussian distributed.  For the GLS estimators, we can conclude from the latter that the residuals $\delta \msd_{i}$ are also normal distributed because Eq.~\eqref{eq:residuals} is linear, which, in turn, justifies the use of the quality factor.

\subsection{MD simulation and data analysis for ubiquitin}\label{subsec:ubiquitin-diffusion-coefficient}

\begin{figure}[t!]
\begin{center}
\includegraphics{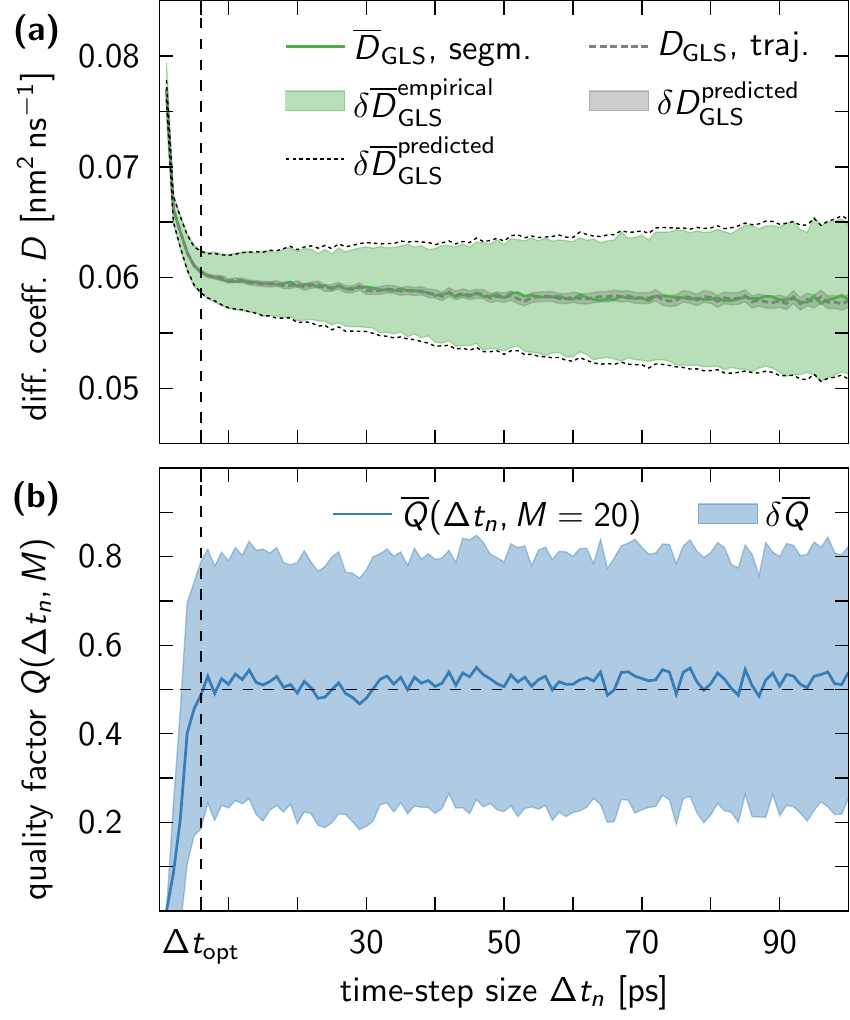}
\caption{Translational diffusion coefficient estimation and quality factor analysis for a single ubiquitin in aqueous solution with $M = 20$ fixed.  (a)~Splitting the trajectory up into 150 segments of equal length allowed us to calculate the sample average [Eq.~\eqref{eq:D-mean}] via the GLS estimators [Eqs.~\eqref{eqs:gls-estimators}, solid green line] and the associated sample standard deviation [Eq.~\eqref{eq:D-std}, green shaded area].  For comparison, the trajectory was also analyzed as a whole (dashed gray line), where the corresponding uncertainty was determined via Eq.~\eqref{eq:D-uncertainty-estimate} (gray shaded area).  Applying Eq.~\eqref{eq:D-uncertainty-estimate} to the sample average $\smash{\overline{D}_{\text{GLS}}}$ resulted in the dotted black lines.  (b)~Analogous to the TIP4P-D water data (see Fig.~\ref{fig:diffusion_coefficient_water}b), the lower bound $\Delta t_{\text{opt}}$ was read off the quality factor plot, giving $\Delta t_{\text{opt}} = \SI{6}{\pico \second}$.  However, unlike for water, the average quality factor (solid blue line) is consistently above $1/2$ for $\smash{\Delta t_{n}} > \smash{\Delta t_{\text{opt}}}$, which is indicative of the trajectory segments being too short (see Fig.~\ref{fig:quality_factor_short_time_series}).  The shaded area represents the uncertainty of our estimate, which was determined from the sample standard deviation.  }
\label{fig:diffusion_coefficient_ubiquitin}
\end{center}
\end{figure}

We tested our framework in a more biological context by simulating a \SI{2}{\micro \second} trajectory of a single ubiquitin molecule (PDB identification code: 1ubq\cite{Vijay-KumarBugg1987}) using the Amber99SB*-ILDN-Q force field\cite{Lindorff-LarsenPiana2010, BestHummer2009, HornakAbel2006, BestdeSancho2012} in a cubic simulation box with periodic boundary conditions and an approximate edge length of \SI{7.5}{\nano \meter}.  The protein was solvated by 13347 TIP4P-D water molecules at a concentration of $\sim$\SI{150}{\milli \molar} NaCl.\cite{JoungCheatham2008}  The equilibration and production runs were otherwise performed in the same manner as described in Sec.~\ref{subsec:TIP4P-D-simulation-setup}.  

In contrast to pure TIP4P-D water, we had only a single trajectory for the ubiquitin protein, which we analyzed in two ways:  First, we applied the GLS estimators [Eqs.~\eqref{eqs:gls-estimators}] to MSD values computed for the full trajectory at different interval lengths.  Then, we split the trajectory into 150 segments of equal length and treated each segment as an individual trajectory in the data analysis, thus allowing for a similar data analysis as performed on the water simulation data.  In both cases, $M=20$ MSD values were considered for the fits.  The results of both approaches are presented in Fig.~\ref{fig:diffusion_coefficient_ubiquitin}a, where we compare the single-trajectory diffusion coefficient estimate $\smash{D_{\text{GLS}}}$ to a sample average $\smash{\overline{D}_{\text{GLS}}}$ over the above-mentioned segments.  The two estimates coincide almost perfectly and possible deviations are contained within the uncertainty interval of $\smash{D_{\text{GLS}}}$.  

\begin{figure}[t!]
\begin{center}
\includegraphics{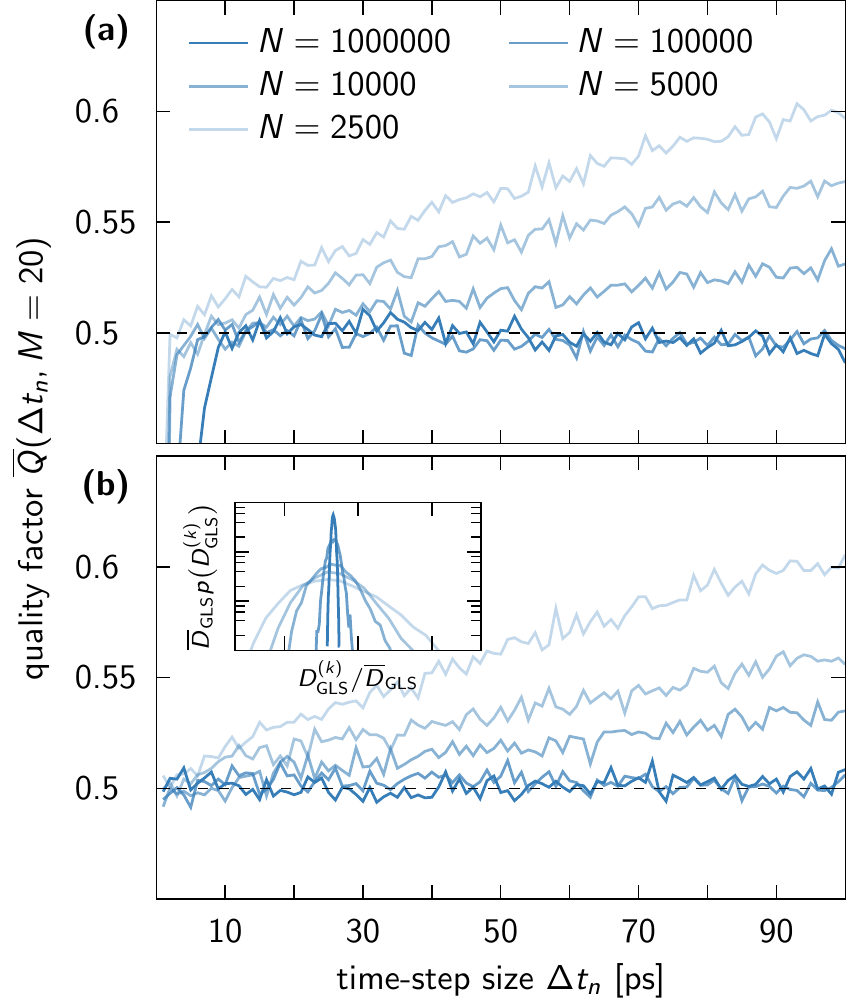}
\caption{For short time series, the average quality factor $\smash{\overline{Q}}$ deviates from $1/2$ as a result of non-Gaussian residuals.  (a)~$\smash{\overline{Q}}$ as function of the recording-time interval length $\Delta t_{n}$ for the TIP4P-D water time series of Fig.~\ref{fig:diffusion_coefficient_water} truncated to different lengths $N$.  (b)~Same as in (a) for ideal diffusive dynamics, generated via Eqs.~\eqref{eqs:Z_X_processes} for $a^{2} = 1/2$ and $\sigma^{2} = 1$.  \emph{Inset:}~Distributions of the corresponding diffusion coefficient estimates for $\Delta t_{n} = \SI{100}{\pico \second}$ on a semi-logarithmic scale.  }
\label{fig:quality_factor_short_time_series}
\end{center}
\end{figure}

\begin{figure*}[t!]
\begin{center}
\includegraphics{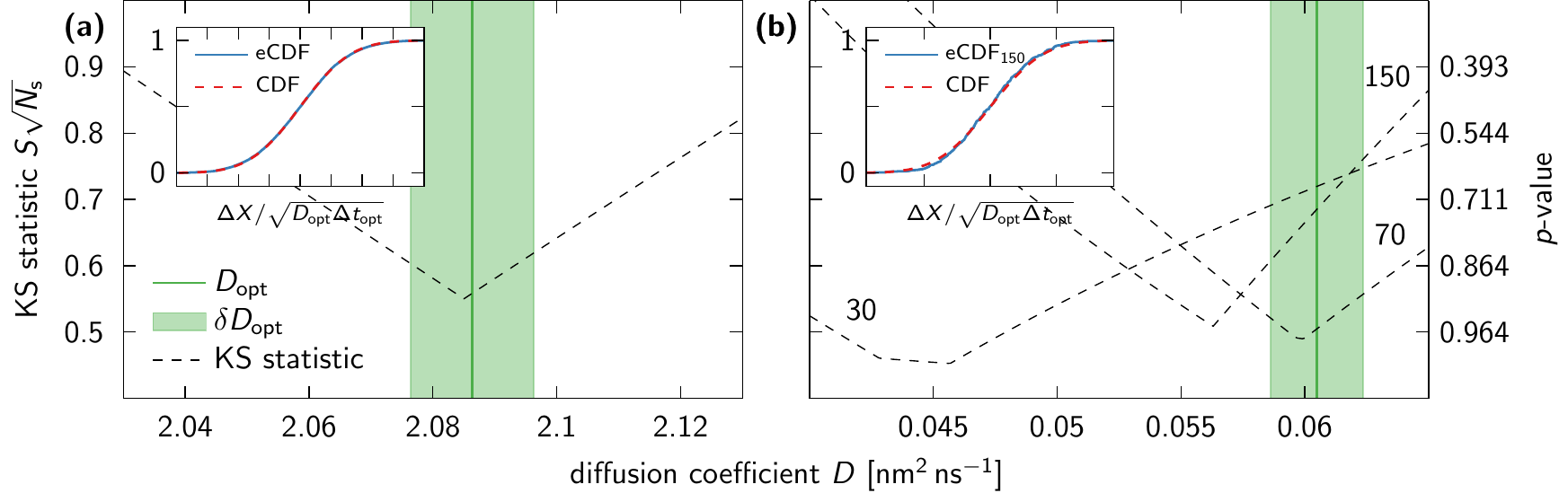}
\caption{Kolmogorov-Smirnov test to verify whether $D_{\text{opt}} = \smash{\overline{D}_{\text{GLS}}}$ obtained for $\Delta t_{n} = \Delta t_{\text{opt}}$ is consistent with the observed long-time dynamics.  (a)~The optimal diffusion coefficient estimate $D_{\text{opt}} = \SI{2.086 \pm 0.010}{\nano \meter \squared \per \nano \second}$ for TIP4P-D water (solid green line), which was read off the quality factor (see Fig.~\ref{fig:diffusion_coefficient_water}b), coincides near perfectly with the minimum of the scaled KS statistic [Eq.~\eqref{eq:kolmogorov-smirnov-statistic}, dashed black line].  Here, $\smash{a^{2}(\Delta t_{\text{opt}})} = \smash{\overline{a}^{2}_{\text{GLS}} \vert_{\Delta t_{n} = \Delta t_{\text{opt}}}}$ was kept constant for all $D$, and $\Ns = 3 \times 4139$.  The shaded area represents one sample standard deviation of uncertainty.  \emph{Inset:}~Empirical (solid blue line) and cumulative distribution function [Eq.~\eqref{eq:cdf}, dashed red line] of the relative endpoints $\smash{\{ \Delta X^{(k)} \}_{k=1,2,\dots,4139}^{d \in \{ x,y,z \}}}$, where the latter was evaluated using $D_{\text{opt}}$.  (b)~Same as in (a) for $N_{\text{seg}} = 30$, $70$ and $150$ ubiquitin trajectory segments using $D_{\text{opt}} = \SI{6.0 \pm 0.2 e-2}{\nano \meter \squared \per \nano \second}$ and $\Ns = 3 \times N_{\text{seg}}$.  Although our estimate $D_{\text{opt}}$ is too high compared to the value that minimizes the KS statistic for $N_{\text{seg}} = 30$ and $150$, we cannot reject it on the basis of the KS test because the corresponding \textit{p}-values are fairly high ($\sim$0.65 and $\sim$0.72, respectively).  This is reflected in the excellent agreement between eCDF and CDF (see inset).  
}
\label{fig:kolmogorov_smirnov_statistic}
\end{center}
\end{figure*}

Yet, unlike for pure TIP4P-D water, the diffusion coefficient estimate for ubiquitin did not converge to a constant value at sufficiently large $\Delta t_{n}$, but instead crept to ever lower values.  This might indicate some underlying non-diffusive dynamics on long time scales, but an analysis of the quality factor, depicted in Fig.~\ref{fig:diffusion_coefficient_ubiquitin}b, showed that it already reached $\smash{\overline{Q}} = 1/2$ at $\Delta t_{\text{opt}} = \SI{6}{\pico \second}$ and then remained essentially flat, thus implying a good fit to our diffusion model.  At the optimal interval length $\Delta t_{\text{opt}}$, the noise parameter for ubiquitin was estimated as $\smash{\overline{a}_{\text{GLS}}^{2}} = \SI{1.9(6)e-4}{\nano \meter \squared}$, corresponding to a signal-to-noise ratio of $\sigma^{2}/a^{2} \approx 11$.  A closer inspection revealed that $\overline{Q}$ consistently took values slightly greater than $1/2$, indicating that we overestimated $\smash{\delta \overline{D}_{\text{GLS}}^{\text{prediction}}}$ in this regime.  This can also be seen in Fig.~\ref{fig:diffusion_coefficient_ubiquitin}a, where the black dotted lines [Eq.~\eqref{eq:D-uncertainty-estimate}] clearly envelop the green shaded area [Eq.~\eqref{eq:D-std}].  The elevated $\smash{\overline{Q}}$-values beyond $\Delta t_{\text{opt}}$ are due to the trajectory segments being too short to produce Gaussian-distributed estimates for $\smash{D^{(k)}_{\text{GLS}}}$.  We also observed this effect for synthetic data generated via Eqs.~\eqref{eqs:Z_X_processes}, as well as our TIP4P-D water data when we drastically shortened the underlying time series.  The effect is demonstrated in Fig.~\ref{fig:quality_factor_short_time_series}, where we plot the average quality factor computed from time series of different lengths.  For short time series, the quality factor increases with the interval length both in TIP4P-D water data and in synthetic data for an ideal diffusion process.  As shown in the inset of Fig.~\ref{fig:quality_factor_short_time_series}b, this is associated with deviations from Gaussian statistics for small sample sizes.  Furthermore, in the case of the TIP4P-D water data, the lower bound $\Delta t_{\text{opt}}$ also seems affected by the shortening of the time series.  This casts doubt on our initial decision to set $\Delta t_{\text{opt}} = \SI{6}{\pico\second}$ for ubiquitin, because the optimal value is probably a few picoseconds higher.

\subsection{Verifying short-time predictions on long time scales}\label{sec:verifying-predictions}

Another indication that $\Delta t_{\text{opt}}$ might be too low for ubiquitin is given by the fact that the uncertainty bounds at $\Delta t_{\text{opt}} = \SI{6}{\pico \second}$ from the single-trajectory analysis do not account for all diffusion coefficient estimates obtained at longer recording-time intervals.  To test whether this discrepancy is due to insufficient statistics or because our model is not entirely adequate to describe the translational dynamics of ubiquitin, we turned to the KS statistic [Eq.~\eqref{eq:kolmogorov-smirnov-statistic}], which we evaluated for our water data and the segmented ubiquitin trajectory, respectively.  Figure~\ref{fig:kolmogorov_smirnov_statistic}a presents the KS test results for TIP4P-D water, which, unsurprisingly, confirm that the system dynamics remains diffusive on long time scales and is well characterized by the optimal diffusion coefficient $D_{\text{opt}} \equiv D(\Delta t_{\text{opt}}) = \SI{2.086 \pm 0.010}{\nano \meter \squared \per \nano \second}$.  For ubiquitin, we observed a significant discrepancy between the global minimum of $S \sqrt{\Ns}$ and $D_{\text{opt}}$, as depicted in Fig.~\ref{fig:kolmogorov_smirnov_statistic}b.  Yet, due to the large \textit{p}-value associated with $D_{\text{opt}}$, we could not reject the null hypothesis that the long-time dynamics of ubiquitin is diffusive with $D \approx D_{\text{opt}} = \SI{0.060 \pm 0.002}{\nano \meter \squared \per \nano \second}$.  According to the KS test, the value of $D_{\text{opt}}$ estimated from the 150 segments, each of length \SI{13.3}{\nano \second}, is consistent with trajectory data at longer recording-time intervals of \SI{28.5}{\nano \second} and \SI{66.7}{\nano \second}, with \textit{p}-values exceeding 0.5 (see Fig.~\ref{fig:kolmogorov_smirnov_statistic}b).

\section{Conclusions}\label{sec:conclusions}

We have proposed a robust framework to extract reliable self-diffusion coefficients and their uncertainties from molecular dynamics simulation trajectories.  We fit a diffusion process to the observed MSD curves using GLS estimators [Eqs.~\eqref{eqs:gls-estimators}], which account for strong correlations in the errors of MSD values at different time lags, and generally outperform other estimators at high and low signal-to-noise ratios.  We allow for possible non-diffusive dynamics at short times by including an instantaneous Gaussian spread in the diffusion process.  By sub-sampling the time series, we identify remaining deviations from regular diffusion at intermediate times.  We extend the recording-time interval until the sub-sampled trajectories become statistically consistent with our diffusion process.  The quality factor $Q$ [Eq.~\eqref{eq:quality-factor}] used to quantify the fit consistency arises naturally in the context of GLS fitting procedures and provides a measure on whether the data are overfitted or underfitted.  In this way, we obtain an estimate of the translational diffusion coefficient that optimally trades off possible systematic errors from non-diffusive dynamics at short times and statistical errors from increasing uncertainties in the MSD values at longer times.  

Our framework can be readily applied to either a single trajectory or a set of trajectories with the help of our Python
data analysis script.\cite{PythonScript}  In the former case, the full trajectory must be split into multiple segments of equal length to perform the quality factor analysis.  As a proof of principle and to demonstrate its use in practice, we applied our framework to molecular dynamics simulation data.  We estimated the translational diffusion coefficient of TIP4P-D water from an ensemble of 4139 water molecules and compared our results to literature values.  We then went on to a system with much sparser statistics, namely a single ubiquitin molecule solvated in TIP4P-D water, where we performed a single-trajectory analysis on the full ubiquitin trajectory and compared the results to statistics obtained by analyzing multiple segments of said trajectory.  We found that both approaches give almost identical diffusion coefficient estimates, albeit with very different uncertainties.  An inspection of the quality factor for the segments revealed that our predictions for the uncertainty were adequate, but systematically a bit too high due to the segments being too short.  With the help of a Kolmogorov-Smirnov test, we confirmed that our optimal diffusion coefficient estimates correctly predict the long-time dynamics observed in our trajectories.  In this way, we effectively ruled out possible anomalous diffusion of the protein ubiquitin on intermediate and long time scales.  Finally, our framework allowed us to identify systematic errors caused by well-established software packages for unwrapping particle trajectories from simulations at constant pressure, as detailed in the companion paper, Ref.~\onlinecite{vonBuelowBullerjahn2020}.  

All in all, our framework supplies the practitioner with an optimal diffusion coefficient estimate and an associated uncertainty with high precision.  It furthermore provides evidence on how to optimize the quality of the fit by sub-sampling, thereby trading off systematic and statistical uncertainties, and whether the predicted overall uncertainty is of appropriate size or not.

\section*{Data availability}
The data that support the findings of this study are available from the corresponding author upon reasonable request.

\begin{acknowledgments}

We thank Attila Szabo for insightful comments on the manuscript and discussions.  This research was supported by the Max Planck Society (J.T.B., S.v.B. and G.H.) and the Human Frontier Science Program RGP0026/2017 (S.v.B. and G.H.).  

\end{acknowledgments}

\begin{appendix}

\section{OLS-estimator variances}\label{app:OLS-variances}

The variances of the OLS estimators [Eqs.~\eqref{eqs:ols-estimators}] are computed in a straight-forward fashion, giving
\begin{align*}
\var (a_{\text{OLS}}^{2}) & = \avg{(a_{\text{OLS}}^{2})^{2}} - \avg{a_{\text{OLS}}^{2}}^{2}
\\
& = \frac{\beta^{2} \avg{\gamma^{2}} - 2 \alpha \beta \avg{\gamma \delta} + \alpha^{2} \avg{\delta^{2}}}{(M \beta - \alpha^{2})^{2}} - a^{4} \, ,
\\
\var (\sigma_{\text{OLS}}^{2}) & = \frac{M^{2} \avg{\delta^{2}} - 2 M \alpha \avg{\gamma \delta} + \alpha^{2} \avg{\gamma^{2}}}{(M \beta - \alpha^{2})^{2}} - \sigma^{4} \, .  
\end{align*}
where we can make use of the relation $\avg{\msd_{i}\msd_{j}} = \Sigma_{i,j} + (a^{2} + i \sigma^{2})(a^{2} + j \sigma^{2})$ to rewrite the ensemble averages appearing in the above expressions as follows,
\begin{align*}
\avg{\gamma^{2}} & = \sum_{i,j=1}^{M} \Sigma_{i,j} + (a^{2} M + \sigma^{2} \alpha)^{2} \, ,
\\
\avg{\gamma \delta} & = \sum_{i,j=1}^{M} i \, \Sigma_{i,j} + a^{4} M \alpha + a^{2} \sigma^{2} (M \beta + \alpha^{2}) + \sigma^{4} \alpha \beta \, ,
\\
\avg{\delta^{2}} & = \sum_{i,j=1}^{M} i \, j \, \Sigma_{i,j} + (a^{2} \alpha + \sigma^{2} \beta)^{2} \, .  
\end{align*}
Here, $\Sigma_{i,j}$ denotes the covariance of the $\msd_{i}$, whose functional form is explicitly derived in the next section and given by Eqs.~\eqref{eqs:covariance_matrices}.  After some algebraic manipulations, we arrive at the final result
\begin{subequations}\label{eqs:ols-variances}
\begin{align}
\var (a_{\text{OLS}}^{2})
& = \sum_{i,j=1}^{M} \frac{(i \alpha - \beta) (j \alpha - \beta)}{(M \beta - \alpha^{2})^{2}} \Sigma_{i,j} (a_{\text{OLS}}^{2},\sigma_{\text{OLS}}^{2}) \, ,
\\
\label{eq:ols-sigma-variance}
\var (\sigma_{\text{OLS}}^{2}) & = \sum_{i,j=1}^{M} \frac{(i M - \alpha) (j M - \alpha)}{(M \beta - \alpha^{2})^{2}} \Sigma_{i,j} (a_{\text{OLS}}^{2},\sigma_{\text{OLS}}^{2}) \, .  
\end{align}
\end{subequations}
For $\sigma^{2}/a^{2} > 1$, the precision of the OLS estimators deteriorates with increasing $M$, as seen in the limit $M \ll N$ with $a^{2} \equiv 0$, where
\begin{align*}
\var (\sigma_{\text{OLS}}^{2}) 
& = \frac{2 \sigma^{4}}{3 N} \sum_{i,j=1}^{M} \frac{(i M - \alpha) (j M - \alpha)}{(M \beta - \alpha^{2})^{2}} \min(i,j)
\\
& \mathrel{\phantom{\sim}} \times \big[ 1 + 3 i j - \min(i,j)^{2} \big] + \mathcal{O}(N^{-2})
\\
& = \frac{2 \sigma^{4}}{3 N} \bigg[ \frac{78 M}{35} + 3 + \mathcal{O} (M^{-1}) \bigg] + \mathcal{O}(N^{-2}) \, .  
\end{align*}
Including more data points therefore worsens the OLS estimate of the diffusion coefficient.  

By contrast, at low signal-to-noise ratios $\sigma^{2}/a^{2}$ other terms dominate the covariance, thus resulting in an uncertainty comparable to the GLS estimators for $M \ll N$ (see Fig.~\ref{fig:variance_comparison}).

\section{MSD covariance matrix}\label{app:derivation-1}

While the mean of $\msd_{i}$ is trivially given by
\begin{equation*}
\avg{\msd_{i}} = \sum_{n=0}^{N-i} \frac{\avg{(X_{n+i} - X_{n})^{2}}}{N-i+1} = a^{2} + i \sigma^{2} \, ,
\end{equation*}
the calculation of its second moment,
\begin{equation*}
\avg{\msd_{i} \msd_{j}} = \sum_{n=0}^{N-i} \sum_{m=0}^{N-j} \frac{\avg{\delta X_{n,i}^{2} \delta X_{m,j}^{2}}}{(N-i+1)(N-j+1)}
\end{equation*}
with the shorthand notation $\delta Y_{n,i} = Y_{n+i} - Y_{n}$, is more involved.  It essentially reduces to a sum of moments of the form $\avg{A_{i} B_{j} C_{k} D_{l}}$, where $A, B, C, D \in \{ Z, X, R, S \}$, which evaluate to
\begin{align}\label{eq:fourth-moment}
\avg{A_{i} B_{j} C_{k} D_{l}}
\notag
& = \cov(A_{i}, B_{j}) \cov(C_{k}, D_{l})
\\
\notag
& \mathrel{\phantom{=}} + \cov(A_{i}, D_{l}) \cov(B_{j}, C_{k})
\\
& \mathrel{\phantom{=}} + \cov(A_{i}, C_{k}) \cov(B_{j}, D_{l})
\end{align}
for normally distributed processes.\cite{Gardiner1985}  Because of
\begin{align*}
\avg{Z_{i} Z_{j} Z_{k} S_{l}} & = \sum_{\mathclap{\text{combinations}}} \cov(Z_{a}, Z_{b}) \cov(Z_{c}, S_{d}) = 0 \, ,
\\
\avg{Z_{i} S_{j} S_{k} S_{l}} & = \sum_{\mathclap{\text{combinations}}} \cov(Z_{a}, S_{b}) \cov(S_{c}, S_{d}) = 0 \, ,
\end{align*}
it becomes apparent that the only non-zero terms of $\avg{\smash{(X_{n+i} - X_{n})^{2}} \smash{(X_{m+j} - X_{m})^{2}}}$, when expanded with respect to $Z$ and $S$, are of even order, \emph{i.e.},
\begin{align*}
\avg{\delta X_{n,i}^{2} \delta X_{m,j}^{2}}
& = \avg{\delta Z_{n,i}^{2} \delta Z_{m,j}^{2}} + \frac{a^{2}}{2} \avg{\delta Z_{n,i}^{2}\delta S_{m,j}^{2}}
\\
& \mathrel{\phantom{=}} + 2 a^{2} \avg{\delta Z_{n,i} \delta S_{n,i} \delta Z_{m,j} \delta S_{m,j}}
\\
& \mathrel{\phantom{=}} + \frac{a^{2}}{2} \avg{\delta S_{n,i}^{2}\delta Z_{m,j}^{2}} + \frac{a^{4}}{4} \avg{\delta S_{n,i}^{2}\delta S_{m,j}^{2}} \, .  
\end{align*}
We can already identify
\begin{equation}\label{eq:a=0_covariance_matrix}
\Sigma_{i,j} (0,\sigma^{2}) = \sum_{n=0}^{N-i} \sum_{m=0}^{N-j} \frac{\avg{\delta Z_{n,i}^{2} \delta Z_{m,j}^{2}}}{(N-i+1)(N-j+1)} - i j \sigma^{4} \, ,
\end{equation}
which we will come back to later, and use linear combinations of
\begin{align*}
\avg{Z_{i} Z_{j} S_{k} S_{l}} & = \sigma^{2} \min(i,j) \delta_{k,l} \, ,
\\
\avg{S_{i} S_{j} S_{k} S_{l}} & = \avg{R_{i} R_{j} R_{k} R_{l}} = \delta_{i,j} \delta_{k,l} + \delta_{i,l} \delta_{j,k} + \delta_{i,k} \delta_{j,l}
\end{align*}
to calculate the remaining terms, giving
\begin{align*}
\avg{\delta Z_{n,i}^{2}\delta S_{m,j}^{2}} & = 2 i \sigma^{2} \, , 
\\
\avg{\delta Z_{n,i} \delta S_{n,i} \delta Z_{m,j} \delta S_{m,j}} & = \sigma^{2} (\delta_{m,n} - \delta_{m,n+i}
\\
& \mathrel{\phantom{=}}  - \delta_{m+j,n} + \delta_{m+j,n+i})
\\
& \mathrel{\phantom{=}} \times [ \min(m,n) - \min(m,n+i)
\\
& \mathrel{\phantom{=}} - \min(m+j,n)
\\
& \mathrel{\phantom{=}} + \min(m+j,n+i) ] \, ,
\\
\avg{\delta S_{n,i}^{2}\delta Z_{m,j}^{2}} & = 2 j \sigma^{2} \, , 
\\
\avg{\delta S_{n,i}^{2}\delta S_{m,j}^{2}} & = 2 [ 2 + (1 + 2 \delta_{i,j}) \delta_{m,n} + \delta_{m,n+i}
\\
& \mathrel{\phantom{=}} + \delta_{m+j,n} + \delta_{m+j,n+i} ] \, .  
\end{align*}
Substituting all of these into our expression for the covariance matrix results in Eq.~\eqref{eq:general_covariance_matrix}.  

For the special case of $a \equiv 0$, our covariance matrix reduces to Eq.~\eqref{eq:a=0_covariance_matrix}, where the ensemble average can be rewritten as follows,
\begin{align*}
\avg{\delta Z_{n,i}^{2}\delta Z_{m,j}^{2}}
& \mathop{=}^{\phantom{\eqref{eq:fourth-moment}}} \sigma^{4} \Bigg\langle \Bigg[ \sum_{a=n}^{n+i-1} R_{a} \Bigg]^{2} \Bigg[ \sum_{b=m}^{m+j-1} R_{b} \Bigg]^{2} \Bigg\rangle
\\
& \mathop{=}^{\eqref{eq:fourth-moment}} \sigma^{4} \Bigg( i j + 2 \Bigg[ \sum_{a=n}^{n+i-1} \sum_{b=m}^{m+j-1} \delta_{a,b} \Bigg]^{2} \Bigg) \, .  
\end{align*}
The $\sigma^{4} i j$-term conveniently cancels with $\avg{\msd_{i}} \avg{\msd_{j}}$ and the summation over multiple indices can be reduced to
\begin{align*}
\sum_{n=0}^{N-i} & \sum_{m=0}^{N-j} \Bigg[ \sum_{a=n}^{n+i-1} \sum_{b=m}^{m+j-1} \delta_{a,b} \Bigg]^{2}
\\
& = \frac{1}{6} \big( 2 \min(i,j) [N + 1 - \max(i,j)]
\\
& \mathrel{\phantom{=}} \times \big[ 1 + 3 i j - \min(i,j)^{2} \big]
\\
& \mathrel{\phantom{=}} + \min(i,j)^{2} - \min(i,j)^{4} + \Theta(i+j-N-2)
\\
& \mathrel{\phantom{=}} \times \big[ (N+1-i-j)^{4} - (N+1-i-j)^{2} \big] \big)
\end{align*}
using geometric reasoning.  This finally gives rise to Eq.~\eqref{eq:specific_covariance_matrix} of the main text.

\section{Iterative algorithm for the GLS estimators}\label{app:algorithm}

Starting from a candidate solution $(\smash{a_{k}^{2}},\smash{\sigma_{k}^{2}})$, we can evaluate the covariance matrix [Eq.~\eqref{eq:general_covariance_matrix}] and use it to construct the following auxiliary values,
\begin{align*}
\kappa_{k} & = \sum_{i,j = 1}^{M} \Sigma_{i,j}^{-1}(a_{k}^{2},\sigma_{k}^{2}) \, , 
\\
\notag
\lambda_{k} & = \sum_{i,j = 1}^{M} i \, \Sigma_{i,j}^{-1}(a_{k}^{2},\sigma_{k}^{2}) \, , 
\\
\notag
\mu_{k} & = \sum_{i,j = 1}^{M} i \, j \, \Sigma_{i,j}^{-1}(a_{k}^{2},\sigma_{k}^{2}) \, , 
\\
\notag
\nu_{k} & = \sum_{i,j = 1}^{M} \msd_{i} \, \Sigma_{i,j}^{-1}(a_{k}^{2},\sigma_{k}^{2}) \, , 
\\
\notag
\xi_{k} & = \sum_{i,j = 1}^{M} i \, \msd_{j} \, \Sigma_{i,j}^{-1}(a_{k}^{2},\sigma_{k}^{2}) \, .   
\end{align*}
Our candidate solution can then be iteratively updated according to
\begin{align*}
a_{k+1}^{2} & = \frac{\mu_{k} \nu_{k} - \lambda_{k} \xi_{k}}{\kappa_{k} \mu_{k} - \lambda_{k}^{2}} \, , \qquad \sigma_{k+1}^{2} = \frac{\kappa_{k} \xi_{k} - \lambda_{k} \nu_{k}}{\kappa_{k} \mu_{k} - \lambda_{k}^{2}} \, ,   
\end{align*}
where the procedure is stopped either when the following inequality is satisfied for some tolerance threshold $\varepsilon$,
\begin{equation*}
\big( a_{k+1}^{2} - a_{k}^{2} \big)^{2} + \big( \sigma_{k+1}^{2} - \sigma_{k}^{2} \big)^{2} < \varepsilon \, , 
\end{equation*}
or a certain number of iteration steps are completed.  As an initial solution, we choose 
\begin{align*}
(a_{0}^{2},\sigma_{0}^{2})
& = (a_{M=2}^{2},\sigma_{M=2}^{2})
\\
& \equiv (2 \msd_{1} - \msd_{2}, - \msd_{1} + \msd_{2}) \, , 
\end{align*}
which is also returned if the algorithm fails to converge.  Although this is rarely the case for $1 < M \ll N$, the above described procedure is susceptible to considerable numerical errors in the limit $M \to N \gg 1$, where the covariance matrix becomes ill-conditioned.

\section{Asymptotics of the GLS estimators}\label{app:derivation-2}

According to Eq.~\eqref{eq:general_covariance_matrix}, we have the identity
\begin{align*}
\Sigma_{1,i}(a^{2},\sigma^{2})
& = \frac{2 i}{N} \sigma^{4} + \frac{a^{4} (2 + \delta_{1,i}) + 4 a^{2} \sigma^{2}}{N}
\\
& \mathrel{\phantom{=}} - \frac{a^{4}}{N (N - i + 1)} \, ,
\end{align*}
from which the relation
\begin{align*}
b_{1}
& = \frac{2}{N} \sigma^{4} \sum_{i,j=1}^{M} i \, b_{j} \, \Sigma_{i,j}^{-1}(a^{2},\sigma^{2})
\\
& \mathrel{\phantom{=}} + \frac{2 a^{4} + 4 a^{2} \sigma^{2}}{N} \sum_{i,j=1}^{M} b_{j} \, \Sigma_{i,j}^{-1}(a^{2},\sigma^{2})
\\
& \mathrel{\phantom{=}} + \frac{a^{4}}{N} \sum_{i,j=1}^{M} \delta_{1,i} \, b_{j} \, \Sigma_{i,j}^{-1}(a^{2},\sigma^{2})
\\
& \mathrel{\phantom{=}} - \sum_{i,j=1}^{M} \frac{b_{j} a^{4}}{N (N - i + 1)} \Sigma_{i,j}^{-1}(a^{2},\sigma^{2})
\end{align*}
follows for arbitrary $b_{i}$.  For large signal-to-noise ratios, we thus have
\begin{align}\label{eq:identity}
\notag
\sum_{i,j=1}^{M} i \, b_{j} \, \Sigma_{i,j}^{-1}(a^{2},\sigma^{2}) & = \frac{N}{2 \sigma^{4}} b_{1} - \frac{2 a^{2}}{\sigma^{2}} \sum_{i,j=1}^{M} b_{j} \, \Sigma_{i,j}^{-1}(a^{2},\sigma^{2})
\\
& \mathrel{\phantom{=}} + \mathcal{O} \big( [ \sigma^{2} / a^{2} ]^{-2} \big) \, ,
\end{align}
which can be used to derive Eqs.~\eqref{eq:asymptotic-gls-estimators}, because
\begin{align*}
\kappa \mu - \lambda^{2}
& = \frac{N}{2 \sigma^{4}} [ \kappa - \lambda ] + \mathcal{O} \big( [ \sigma^{2} / a^{2} ]^{-2} \big) \, , 
\\
\mu \nu - \lambda \xi & = \frac{N}{2 \sigma^{4}} [ \nu - \msd_{1} \lambda ] + \mathcal{O} \big( [ \sigma^{2} / a^{2} ]^{-2} \big) \, , 
\\
\kappa \xi - \lambda \nu & = \frac{N}{2 \sigma^{4}} [ \msd_{1} \kappa - \nu ] + \mathcal{O} \big( [ \sigma^{2} / a^{2} ]^{-2} \big) \, , 
\end{align*}
must hold.  Finally, Eq.~\eqref{eq:gls-variance_a=0} follows from Eq.~\eqref{eq:identity} for $a \equiv 0$ and $b_{j} = j$.

\end{appendix}

\end{document}